\documentstyle[11pt,aaspp4]{article}

\def\etal{et al.\ }
\def\ie{{i.e.\hskip 3pt}}
\def\eg{{e.g.\hskip 3pt}}

\newcommand{\lya}{Ly$\alpha$\ }
\newcommand{\lyb}{Ly$\beta$\ }
\newcommand{\nh}{N_{\rm HI}}
\newcommand{\kms}{\;{\rm km}\,{\rm s}^{-1}}

\newcommand\cdunits{{\rm cm}^{-2}}

\slugcomment{Submitted to ApJ}

\lefthead{Dav\'e et al.}
\righthead{Metallicity of the \lya forest from OVI}
\singlespace

\begin{document}

\title{Constraining the Metallicity of the Low Density Lyman-Alpha Forest
  Using OVI Absorption}

\author{Romeel Dav\'e, Uffe Hellsten and Lars Hernquist}
\affil{Astronomy Department, University of California,
    Santa Cruz, CA 95064}

\author{Neal Katz}
\affil{Astronomy Department, University of Massachusetts, Amherst, MA 01003}

\and

\author{David H. Weinberg}
\affil{Astronomy Department, Ohio State University, Columbus, OH 43210}

\begin{abstract}

We present a systematic search for OVI(1032\AA,1037\AA) absorption in a
Keck HIRES spectrum of the $z=3.62$ quasar Q1422+231, with the goal of
constraining the metallicity and ionization state of the low density
intergalactic medium (IGM).  Comparison of CIV absorption measurements
to models of the \lya forest based on cosmological simulations shows
that absorbers with $\nh \ga 10^{14.5}\cdunits$ have a mean carbon
abundance [C/H]~$\approx -2.5$, assuming a metagalactic photoionizing
background with the spectral shape predicted by Haardt \& Madau (1996,
HM).  In these models, lower column density absorption arises in lower
density gas where most CIV is photoionized to CV.  Therefore, OVI
should be the most sensitive tracer of metallicity in \lya absorbers
with $\nh \la 10^{14.5}\cdunits$.  OVI lines lie at wavelengths heavily
contaminated by Lyman series absorption, so we interpret the search
results by comparing to carefully constructed, mock Q1422 spectra drawn
from a hydrodynamic simulation of a $\Lambda$-dominated cold dark
matter model.

A search for deep, narrow absorption features yields only a few
candidate OVI lines in the spectrum of Q1422.  HI absorption blankets
the position of the doublet companion line in each case, and the total
number of narrow lines is statistically consistent with that in
zero-metallicity artificial spectra.  Artificial spectra generated with
the HM background and [O/H]~$\ga -2.5$ predict too many narrow lines
and are statistically inconsistent with the data.  We also search for
OVI associated with CIV systems, using the optical depth ratio
technique of Songaila (1998).  With this method we {\it do} find
significant OVI absorption; matching the data requires [O/C]~$\approx
+0.5$ and corresponding [O/H]~$\approx -2.0$.  Taken together, the
narrow line and optical depth ratio results imply that (a) the
metallicity in the low density regions of the IGM is at least a factor
of three below that in the overdense regions where CIV absorption is
detectable, and (b) oxygen is overabundant in the CIV regions,
consistent with the predictions of Type II supernova enrichment models
and the observed abundance pattern in old halo stars.

The photoionizing background spectrum would be truncated above 4~Ry in
regions that have not undergone helium reionization
(HeII$\longrightarrow$HeIII), and in this case matching the Q1422 data
requires lower [C/H] but higher [O/H]. Taking [O/C]$\approx +1$ as the
maximum plausible overabundance of oxygen, we conclude that helium must
have been reionized through at least 50\% of the volume from $z \sim 3
- 3.6$.

\end{abstract}


\section{Introduction}

The Lyman alpha (Ly$\alpha$) ``forest" (\cite{lyn71}; \cite{sar80}) of
spectral features caused by HI absorption along the line of sight to a
quasar probes the state of the intergalactic medium over a wide range
of physical conditions.  During the past few years, high-precision
observations made using the HIRES spectrograph (\cite{vog94}) on the
10m Keck telescope have quantified the statistics of these low column
density absorbers to unprecedented accuracy (e.g., \cite{hu95};
\cite{lu96b}; \cite{kim97}; \cite{kir97}).  During the same time span,
cosmological simulations that incorporate gas dynamics, radiative
cooling, and photoionization have been able to reproduce many of the
observed properties of quasar absorption spectra (\cite{cen94};
\cite{zha95}; \cite{her96}; \cite{mir96}; \cite{dav97a}).  The rapid
progress on theoretical and observational fronts has led to the
emergence of a new paradigm for the origin of the high-redshift ($z\ga
2$) \lya forest, in which most \lya forest lines are produced by
regions of low to moderate overdensity in hierarchically collapsing
structures that are not in dynamical or thermal
equilibrium.  \lya lines of lower column density generally arise in
gas of lower physical density, which has a lower neutral hydrogen fraction
because of the reduced recombination rate.  In this paper, we use a
matched comparison between a cosmological hydrodynamic simulation
and a Keck HIRES spectrum of the quasar Q1422+231 to constrain the
metal abundance of this low density gas.

The recent detection of metal lines associated with \lya forest
absorbers having column densities $\nh \la 10^{15} \cdunits$
(\cite{cow95}; \cite{tyt95}; Songaila \& Cowie 1996, hereafter
\cite{son96}) has provided a new avenue for investigating the
ionization state and enrichment history of the high-redshift
intergalactic medium (IGM).  \cite{son96} showed that 75\% of \lya
absorbers with $\nh > 10^{14.5} \cdunits$ have associated CIV
absorption.  Using simple photoionization models they estimate that the
mean metallicity of these absorbers is between\footnote{We use the
standard notation of brackets to denote the relative abundance, in
logarithm, versus solar.} [C/H]$\sim -2$ and $-3$.  
Studies that use cosmological simulations to model the density
and temperature of the absorbing gas obtain
a better match to the data for a mean
metallicity of [C/H]$\sim -2.5$ with around one dex of scatter,
assuming either a power-law ionizing background (\cite{hae96}) or a
reprocessed quasar ionizing background (Hellsten \etal 1997, hereafter
\cite{hel97}).

The question of how metals came to reside in these intermediate-density
\lya absorbers remains unanswered.  Since \lya absorbers with $\nh
\la 10^{15}\cdunits$ are optically thin and associated with density
peaks of overdensity $\la 20$, it is unlikely that they 
contain star-forming regions that produce {\it in situ}
enrichment.  Thus some transport mechanism must be invoked to explain
the presence of metals in these regions.  One possibility is that the
metals are ejected from nearby, forming galaxies, by some
combination of supernova blowout (\cite{mir97}) and tidal stripping 
(\cite{gne97}; \cite{gne98}). 
Since the enriching proto-galaxies are likely to form more efficiently
in high density environments, these scenarios predict a significant
correlation between metallicity and density (see, e.g., figure~6 
of Gnedin \& Ostriker 1997).
Alternatively, the IGM may
have been enriched at a very early epoch by more ubiquitous
Population III objects (e.g., \cite{hai97}), in which case all \lya
absorbers might be expected to have roughly the same metallicity.
Because CIV lines probe only a small range of HI column densities (and
hence physical densities) with adequate statistics, it is difficult to
distinguish between these enrichment models using only CIV data.
Metal abundances can be studied at higher densities
in Lyman limit systems ($\nh \ga 10^{17}\cdunits$)
and damped \lya systems ($\nh \ga 10^{20}\cdunits$),
but these are objects where {\it in situ} enrichment is likely and where
(especially for Lyman limit systems) uncertain radiative transfer
effects complicate the inference of metal abundances from line strengths.

In this paper, we attempt to extend metallicity constraints
to the low column density ($\nh \la 10^{14.5}\cdunits$),  
and hence low physical density, \lya forest.
Carbon is difficult to detect in this regime
because the low density reduces $N_{C}$ and ionizes more CIV to CV.
However, \cite{lu98} have used composite spectra to attempt to detect
CIV in such systems, and we briefly discuss their results in \S\ref{sec:
disc}.  In this paper, we focus instead on OVI, which is 
expected to be the one detectable metal absorption feature
tracing \lya absorbers with $\nh \la 10^{14.5}\cdunits$ 
because of its high ionization state and
large oscillator strength (Hellsten \etal 1998, hereafter \cite{hel98}).
The difficulty with this approach, and the reason that it has not
been previously attempted, is that the OVI absorption
features lie embedded within the \lya forest, which is quite crowded at
these redshifts.  We overcome this problem by using a line
identification scheme specifically designed to select candidate
OVI features and by using artificial spectra 
extracted from a realistic cosmological simulation to calibrate
the efficiency of OVI detection and the contamination from narrow \lya lines.
With these procedures, we can test whether the low density regions
of the \lya forest are consistent with a uniform metallicity extrapolated
from the CIV data at higher densities.

The low density IGM is highly photoionized by the metagalactic
ultraviolet (UV) background.  For our standard spectral shape,
we assume that the UV background is produced by quasar emission
reprocessed by \lya forest absorption (Haardt \& Madau 1996,
hereafter \cite{haa96}).
We find that if the IGM metallicity is uniform at [C/H]$\sim -2.5$, the
UV background has the spectral shape given by \cite{haa96},
and oxygen has the factor of
three overabundance (relative to solar) predicted by Type II supernova
enrichment models, then OVI should be readily detectable in the
spectrum of Q1422+231. However, our detection algorithm finds very few
candidate OVI lines in the spectrum.  The absence of detectable OVI features 
has several possible interpretations:  
(1) oxygen is not overabundant relative to carbon in the
low density IGM,  
(2) the metallicity of low density
regions as traced by $\nh \la 10^{14.5}\cdunits$ \lya absorption systems
is lower than the metallicity of intermediate-density regions traced by
higher column density absorption, or
(3) there are many fewer high-energy photons capable of
photoionizing OV to OVI than are predicted by the \cite{haa96} ionizing
background.

To distinguish between these interpretations, we apply a second algorithm,
the optical depth ratio technique of Songaila (1998, hereafter
\cite{son98}) designed to detect OVI in regions where significant CIV
absorption is found.  Since we have independently determined [C/H] in these
regions, we can use this technique to discriminate between different
ionization conditions and abundance patterns.  By applying this technique
to the spectrum of Q1422+231 and calibrating the results using artificial
spectra, we find that:
(1) a significant metallicity gradient (declining metallicity with 
declining density) must exist regardless of whether
helium has mostly reionized or not,
(2) if helium has reionized by $z\sim 3.6$, our results are consistent with 
[O/C]~$\approx +0.5$,
and (3) if the epoch of helium reionization does not begin until $z\sim 3$, 
a highly implausible oxygen overabundance of [O/C]~$\ga +2$ is required.

Combining the results from these two analysis techniques, our favored
scenario for the low density IGM at $z\ga 3$ is one in which more than
half of the volume of the universe has helium predominantly reionized
by $z\sim 3.6$, oxygen is overabundant relative to carbon by a
factor $\ga 3$, and spatial regions with overdensities $\sim 10$
have an average metallicity {\it at least} a factor of 3 higher than
regions near the mean baryonic density.

Section~\ref{sec: modeling} describes our cosmological simulation,
the Q1422+231 data, and our procedure for constructing artificial 
absorption spectra.  Section~\ref{sec: OVIsearch} discusses previous
OVI searches and reviews \cite{hel98}'s argument that OVI should
be the most effective tracer of metallicity in \lya absorbers with
$\nh \la 10^{14}\cdunits$.  Section~\ref{sec: civ} examines CIV
absorption at intermediate column densities, repeating the general
arguments of \cite{hel97} and Rauch et al.\ (1997a) but with a much 
closer match between the theoretical and observational analysis procedures.
Section~\ref{sec: search} is the heart of the paper.  It describes
our algorithm for identifying candidate OVI lines, presents the results
of the OVI searches in the real and artificial spectra, and discusses
the properties of candidate OVI absorbers in Q1422+231 and the simulations.
Figures~\ref{fig: auto5}--\ref{fig: OVIsel} demonstrate the paper's
central results.  Section~\ref{sec: assumptions} discusses the 
impact of varying the assumptions of our standard theoretical model.
Section~\ref{sec: pixmet} describes the optical depth ratio technique for
quantifying OVI absorption, shows the results from this technique
applied to Q1422+231 and artificial spectra, and discusses these
results in conjunction with the results from Section~\ref{sec: search}.
Section~\ref{sec: disc} summarizes our conclusions and discusses them
in light of other recent observational and theoretical developments.

\section{Modeling the Lyman Alpha Forest}\label{sec: modeling}

\subsection{Simulation}\label{sec: sims}

We perform a cosmological hydrodynamic simulation of a
$\Lambda$-dominated cold dark matter (LCDM) model with
$\Omega_\Lambda=0.6$, $\Omega_{\rm CDM}=0.3527$, $\Omega_b=0.0473$,
$H_0=65 \kms {\rm Mpc}^{-1}$, inflationary spectral index $n=0.95$, and
rms fluctuation amplitude $\sigma_8=0.8$.  These values yield a
COBE-normalized LCDM model that is in agreement with most current
observational constraints (\cite{mir96}).  We chose the LCDM model
because it reproduces the characteristics of the \lya forest of
Q1422+231 most closely (Dav\'e \etal, in preparation), although the
differences from other popular cosmologies are slight.  We choose the value of the
baryon density $\Omega_b$ to match recent
observations of the deuterium abundance that give $\Omega_b\approx
0.02 h^{-2}$ (\cite{bur98a}; \cite{bur98b}).  Matching the
observed mean transmission in the \lya forest using reasonable
estimates for the intensity of the metagalactic photoionizing background 
also requires a baryon abundance of this order 
(\cite{her96}; \cite{rau97}; \cite{wei97}).

The simulations were performed using PTreeSPH (\cite{dav97b}), a
version of the TreeSPH code (\cite{her89}; \cite{kat96}) implemented on
massively parallel supercomputers.  We use $64^3$ dark matter particles
and $64^3$ gas particles in a periodic cube 
11.111$h^{-1}$~comoving Mpc on a side and a gravitational softening
length of 3$h^{-1}$~comoving kpc (equivalent Plummer softening).
The masses of the dark
matter and gas particles are $9.64\times 10^8 M_\odot$ and $1.37\times 10^8
M_\odot$, respectively.  
The initial fluctuations are a Gaussian random field with the power
spectrum computed using the Hu \& Sugiyama (1996) formulation of
the CDM transfer function.  We include radiative cooling 
for primordial composition gas and photoionization and heating from
a spatially uniform UV background.  We adopt the UV spectral shape
and intensity (versus redshift) from \cite{haa96}, with the
exception that the intrinsic quasar spectrum has been softened from a
power law index of $-1.5$ to $-1.8$, as is now favored by those authors
(Madau, private communication); we shall still refer to this as the
\cite{haa96} spectrum.  We compute the relative abundances of ionic
species assuming equilibrium between ionization and recombination
(see \cite{kat96}); we will
examine possible effects of this assumption in \S\ref{sec: hot}.  We include
a prescription for star formation, but for the results contained in
this paper, this has no significant effect (\cite{wei96}).  We ran from
a starting redshift of $z=49$ to $z=2$.  This simulation took roughly
3400 node-hours on 16 processors of the Cray T3E at the Pittsburgh
Supercomputing Center.

\subsection{Spectrum of Quasar Q1422+231}

In this paper we will be comparing our simulations to only one quasar
spectrum, that of Q1422+231\footnote{Hereafter, we will
refer to Q1422+231 as ``Q1422".} (shown in the upper left panel of
Figure~\ref{fig: qspec}), since that
is the only one currently available to us.  It was generously provided
to us by A.~Songaila and L.~Cowie, and their observations of Q1422
using Keck's HIRES spectrograph are described more fully in \cite{son96}.
Here we
briefly note that the quasar emission redshift is $z=3.62$, 
and the spectral resolution (as provided to us) is
0.06\AA, corresponding to roughly $4.5-3.2\kms$ per pixel from
3900\AA\  to 7200\AA.
The signal-to-noise ratio is roughly $30-40$ per pixel in the region of \lyb
and OVI absorption ($\lambda\la 4800$\AA), $50-60$ per pixel in the pure \lya forest
region, and over 100 per pixel in the metal line region redwards of the
\lya emission peak.  
Due to a Lyman limit system at $z=3.3816$, the spectrum
of Q1422 effectively truncates around 4000\AA.  Thus
$\lambda \sim$4000\AA--4800\AA\  
represents the usable region for our OVI search, corresponding to
$z\sim 2.9-3.6$, with poorer sensitivity towards lower redshifts
because of
the dropping intrinsic flux of the quasar and the degrading response of
HIRES towards the blue.

\subsection{Continuous Artificial Spectra}

To model Q1422 as closely as possible, we have
developed a new method for generating artificial
quasar spectra from a cosmological simulation.  Our
approach produces continuous line-of-sight artificial spectra between
two desired wavelengths, analogous to real quasar spectra, rather than
a large number of small, disjoint spectra at a single redshift as has
been typically done in other studies (e.g., \cite{her96}).  The
advantages of these {\it continuous artificial spectra} are that they
automatically include the redshift evolution intrinsic to any given
model and that the similarity between the artificial spectra and the
real quasar spectrum allows the identical analysis routines to be
applied to both.  This reduces the algorithm-dependent variations of
continuum fitting, noise estimation, and most importantly for this
work, metal line identification.

Generating continuous spectra requires that particle information be
output at frequent intervals during a simulation.  Since we cannot
output all the simulation data every few timesteps owing to disk space
and I/O time constraints, we output only the particle information
required to generate the desired spectra.  We choose the frequency of output
data so that the redshift interval between outputs exactly
equals the length of the box in redshift space, given by
\begin{equation}
\Delta z = L H(z) / c = 3.33 \times 10^{-4} [H(z)/H_0] L,
\end{equation}
where $H(z)$ is the Hubble constant at redshift
$z$, $H_0$ is the Hubble constant today, $L$ is the simulation size in
comoving $h^{-1}$Mpc, and $c$ is the speed of light.  At
each output timestep during the simulation run, we choose six random
lines of sight from within the volume.  We output information (\ie
$m,{\bf x},{\bf v},T,\rho,h_{\rm smooth}$) for all gas particles whose
SPH smoothing volume (i.e., the region of space a gas particle represents)
intersects one of these lines of sight.  At each output,
a new set of six lines of sight is randomly chosen.  We begin outputting
spectra continuously at $z=4.2$; roughly 100 outputs are required
to $z=2$.

We use this particle information together with the assumed ionizing background
interpolated to the output redshift to generate optical depths
along these lines of sight for the following ions: HI, HeI, HeII, CIV,
CII, SiIV, NV, and OVI.  We use CLOUDY 90 (\cite{fer96}) to compute
lookup tables for the fraction of gas in each ionization state at a
given density and temperature for our adopted $J_\nu$, accounting for
both photoionization and collisional ionization, and we use
routines extracted from the TIPSY package (\cite{kat95}) to compute
smoothed gas properties along each line of sight; see \cite{hel97} for
more details.  We assume that the gas is optically thin, which is
appropriate for the \lya forest regime.  
We compute the profiles of optical depth versus observed wavelength (or,
equivalently, versus redshift) individually for each 
output box, then join the contiguous output boxes to
produce continuous spectra for redshifts below $z=4.2$
along each of the six lines of sight.

When producing these continuous spectra, we have to pay special
attention to the regions where the spectra from contiguous
output boxes are joined, because the
physical properties there are discontinuous.  Since
the simulation volume is periodic, we can arbitrarily shift the particle
positions, and hence the computed optical depths, in a periodic sense
within a single output box.  We shift the optical depths so
that the point of lowest HI absorption lies at the ends of the spectra
at any given output.  Thus we join the spectra in regions of very low
optical depth, which for our purposes are uninteresting; they also arise
from voids that have relatively smooth and homogeneous physical
properties.  We smooth these transition regions
over a small interval of five pixels out of a
total of 1000 output pixels (where one pixel in our artificial spectrum
corresponds to $\sim 1.5\kms$ at these redshifts).

With a list of optical depths versus redshift for various ions, we are
now equipped to produce artificial quasar spectra.  We select a quasar
spectrum to simulate (in our case Q1422), specifying its redshift, the
wavelength interval over which we wish to produce a spectrum, its
resolution in wavelength space, and a spatially uniform metallicity.
We assume an abundance pattern typical of
low-metallicity stars and HII regions, presumably enriched by Type II
supernovae: $Z\equiv$~[Fe/H]~=~[C/H], [N/C]~$=-0.7$, [Si/C]~$=+0.4$, and
[O/C]~$=+0.5$; we discuss these ratios further in \S\ref{sec: O/C}.  In
each wavelength bin, we sum the contribution to the optical depth
from all ions with redshifts lower than the quasar redshift.  We consider the
entire Lyman series up to Ly-30, and both doublet components for the
metal species.  This procedure yields a list of ``observed''
optical depths versus wavelength at the specified spectral resolution.

We now add noise and fit a continuum to each artificial spectrum.
Although we know the true location of the continuum in advance, it is
important to perform a continuum fit to maintain equivalence between
the procedures for analyzing observed and artificial spectra.  We have
developed automated routines for estimating the signal-to-noise ratio
on a pixel-by-pixel basis and fitting a continuum to raw quasar data.
We estimate the noise by first identifying small regions of saturation 
and measuring the average detector noise, and then finding regions
of ``continuum" to estimate the average noise from photon statistics,
within 30\AA\  segments of the spectrum.  In the dense \lya
forest it is difficult to find sizable regions where the flux is
near the continuum, so typically the shot noise is 
somewhat overestimated because it includes a contribution from
fluctuating low level absorption.  We
have not corrected for this effect.  Nevertheless, our noise estimates
agree reasonably well with those of \cite{son96}.  We estimate the continuum
by fitting smoothed second-order polynomials to the spectral peaks in
30\AA\  intervals.  While this may appear crude, 
no method can recover the true continuum because even in
the artificial spectra the flux rarely approaches the true continuum level for
$z\ga 3$.  Thus we strive only to implement a method that can be applied
identically to the real and artificial spectra, is similar to
methods traditionally used in observational analyses,
and uses a low order fitting function in order to minimize 
``overfitting'' of true absorption fluctuations.  Because our OVI
search below is based on identification of narrow isolated absorption
features, the details of continuum fitting are not critical
for our current purposes.

We estimate the continuum and noise characteristics of Q1422 using the
above algorithms.  To incorporate noise into the artificial spectra, we
take the detector noise and shot noise as estimated in the
observed spectra and add it directly (with a Gaussian random distribution) to
the intensity of the artificial spectra,  giving the artificial spectra
very similar noise characteristics to the observed spectra.
We continuum fit the artificial spectra, and the resulting data
are output as a single artificial quasar spectrum.

The intensity of the metagalactic photoionizing background 
($J_\nu$) is not tightly constrained by current observations.
We use the intensity predicted by \cite{haa96} when evolving the
simulation, but because $J_\nu$ does not significantly affect the
dynamics in the optically thin, low-to-moderate overdensity regions, we
can vary the intensity of $J_\nu$ {\it a posteriori} to desired levels
and obtain nearly identical results to having redone the entire
simulation with that $J_\nu$ (\cite{wei97}).  To constrain
$J_\nu$, we note that the mean opacity $\langle e^{-\tau} \rangle$
of the \lya forest depends on the parameter combination 
$\Omega_b^2/J_\nu$, since the HI optical depths are proportional to
this combination (for fixed gas temperature).  
After obtaining a noise-added, continuum-fitted artificial
spectrum, we measure its mean transmission, then use that information
to adjust $J_\nu$, and repeat the above procedure to produce a new
quasar spectrum.  We typically iterate several times until the mean
transmission converges to within 1\% of the observed mean transmission
for Q1422; this corresponds to a determination of $J_\nu$ to better
than 2\% (given our adopted $\Omega_b$).  
We perform this process individually for each artificial
quasar spectrum.  For our LCDM model, the average factor by which the
\cite{haa96} intensity must be increased is $1.29 \pm 0.18$,
where the variance is computed over the six artificial spectra.
As a side note, the implied value of $\Omega_b$ for consistency with
the \cite{haa96} intensity itself is $0.0176\pm 0.0011 h^{-2}$
for this model.

Once we have an intensity of $J_\nu$ for each artificial spectrum, we
now recalculate the metal species' optical depths using this new
intensity, because unlike HI absorption, metal absorption does not scale
in a trivial way with $J_\nu$.  Once we calculate the new optical depths,
we again fit a continuum and add noise in the
manner described above.  Therefore, each artificial
spectrum has the correct HI mean transmission, and the metal absorption
is computed assuming a spatially uniform metallicity and ionizing
background with the \cite{haa96} shape and an intensity constrained by
the mean transmission of Q1422.

In Figure~\ref{fig: qspec} (left panels) we show the spectrum of Q1422 (reproduced from \cite{son96}) and an example of
one of our six artificial spectra for Q1422 with metallicity $Z=-2.5$,
with the corresponding continuum
fit and noise level (multiplied by four for visibility) in the \lya
forest region.  
The differences between these two spectra are due to:
(1) emission features in the quasar not present in the artificial
spectra, (2) the downward slope of the observed continuum in the \lya forest 
of Q1422 caused by the intrinsically varying spectral intensity of the quasar
and the declining response of the HIRES spectrograph towards
the blue, and (3) the intrinsic fluctuations in the continuum level of
Q1422.  These features in Q1422 are subsequently normalized out by
fitting a continuum level to the peaks, as shown in the upper right panel
of Figure~\ref{fig: qspec}; the identical
continuum fitting routine is applied to the artificial spectrum for
consistency.  After this procedure, the artificial spectrum contains
all the significant characteristics of the Q1422 spectrum, as shown in the
lower right panel of Figure~\ref{fig: qspec}.  We also fit a continuum
to the spectra redwards of the \lya peak; this procedure is much more
robust since there are plenty of regions of virtually unabsorbed continuum.

There are some features of real quasar spectra that are not reproduced
in these continuous artificial spectra, although we expect that
these differences will not significantly affect our results.  First,
correlations of lines across the boundaries where the segmented spectra
have been joined are not properly reproduced; thus caution must be
taken when using these spectra to measure line-of-sight correlation
functions.  In this paper, we are counting and measuring only isolated
lines, so this problem does not arise.  Second, metal line absorption
from systems at low redshift is not properly taken into account, since
our artificial spectra stop at $z=2$.  In practice, the influence
of interloping metal lines on the mean transmission in the \lya forest region 
is extremely small, so our determination of 
$J_\nu$ is virtually unaffected.  Third, real spectra have intrinsic
small-scale variations in their noise level, e.g., owing to
a fluctuating response across an Echelle order.  The noise level of the
artificial spectra is constant over a $\sim 30$\AA\  scale, so these
small-scale fluctuations are not reproduced.  We do not expect this to
have a significant effect on our conclusions, but in future
work we will model some of these effects in greater detail.

\section{Previous Searches for OVI}\label{sec: OVIsearch}

Previous searches for OVI absorption at high redshifts have had very
limited success.  The only claimed detection of an OVI line at $z\ga 2$ 
is by Kirkman \& Tytler (1997; hereafter \cite{kir98}).
They found a single narrow feature in Q1422 which they identified as
OVI primarily on the basis of associated CIV absorption.
The doublet companion was subsumed by an HI line.

The method of searching for OVI associated with CIV is taken from
low-redshift studies (\cite{lu91}; \cite{bur96}).  However, given a
universe with a homogeneous metallicity and an ionizing background shape
similar to that predicted by \cite{haa96}, 
this strategy is not the optimal way to find OVI.  
The reason for this is illustrated by the Line Observability Index
(LOX) plot in Figure~\ref{fig: LOX}.

The LOX is described more fully in \cite{hel98}; here we outline the
relevant features.  The LOX is calculated using the density-temperature
relation taken from a balance between cosmic expansion and
photoionization heating, which predicts $T\propto \rho^{0.7}$, and the
peak density-column density relation computed from artificial spectra,
\begin{equation}\label{eqn: rho-NHI}
\log n_{\rm H} = -14.7 + \log {\Omega_b h^2\over 0.02} + 0.7\log \nh .
\end{equation}
Using these relations, 
a given HI column density can be associated with a particular
ionization condition, from which the fraction in any ionization state can
be calculated using CLOUDY 90 (\cite{fer96}).
Combined with an assumed (uniform) metallicity, here taken to be
$10^{-2.5}$ solar for CIV and $10^{-2}$ solar for OVI, this analysis predicts
a rest equivalent width $W_{r\lambda}$ for a typical absorption line of
a given species.  By the appropriate choice of additive constants 
(see \cite{hel98}), the LOX is given by 
\begin{equation}\label{eqn: LOX}
{\rm LOX}\approx \log(W_{r\lambda} / 1 {\rm m\AA}),
\end{equation}
where the equation is exact in the limit of weak lines.

Figure~\ref{fig: LOX} shows the LOX of OVI and CIV given the
\cite{haa96} spectrum (solid curves).  The horizontal dot-dashed line
corresponds roughly to the detectability limit in the \lya forest of
Q1422, while the horizontal dotted line reflects the detectability
limit redwards of the \lya emission peak.  The dashed curves will be
addressed in \S\ref{sec: cut}.  This figure is similar to 
Figures~3--5 of \cite{hel98} except that the LOX is now calculated for the
LCDM model used in this paper, which has a higher baryon fraction.
The impact of the higher $\Omega_b$ is minimal, as \cite{hel98}
predicted it would be.

As stated in \cite{hel98}, given the assumed metallicity and ionization
state, CIV will be observable in lines with HI column densities $\nh
\ga 10^{14}\cdunits$, while OVI will be observable to significantly
lower HI column densities, down to almost $10^{13}\cdunits$ (if the 
OVI line is not subsumed by an HI line).  Since the column density
distribution $d^2N/d\nh dz \propto \nh ^{-1.7}$ (\eg \cite{hu95}; \cite{dav98}),
there are many more HI lines at
lower column densities.  This means that in any given quasar spectrum,
if the metallicity is uniform across all HI column densities,
photoionized OVI should be associated with HI lines that
often have no associated CIV absorption.  We demonstrate this
point more quantitatively in \S\ref{sec: Nhist} below.

\cite{kir98} claimed that the OVI detected in their $z=3.3816$ Lyman limit
system in Q1422 must be collisionally ionized, in part by arguing
that few high-energy photons capable of ionizing oxygen to OVI would
exist far from quasars.  They argued that while \cite{haa96} assumed
uniform emissivity of quasars, in reality quasars are discrete sources,
and far from quasars there should be further softening of the radiation
field due to reprocessing by ambient HeII.  They calculated the mean free
path of HeII-ionizing (4~Ry) 
photons to be roughly $5 h^{-1}$Mpc, and from this value obtained a
volume filling factor of only 2\%.

While valid for HeII, the extension of this argument to OVI is less
certain.  Photons capable of ionizing OV to OVI require
energies greater than 114~eV~$\approx$~8.4~Ry.  Since the cross section of
absorption scales as $\nu^{-3}$, the mean free path of an
OV ionizing photon is $(8.4/4)^3 \approx 10$ times that
of an HeII ionizing photon.  Given this additional factor of 
about 1000 in volume, a universe with the observed distribution of
quasars and a \cite{haa96} ionizing background will effectively be
transparent to OV-ionizing photons.  This does not imply that the OVI
absorption system detected by \cite{kir98} is necessarily photoionized ---
it is associated with a high column density HI absorber that might
well be embedded in a collapsed gas halo hot enough to cause collisional
ionization.  Nevertheless, while our assumption of a
uniform ionizing background may be an oversimplification, we do not
expect that its spectrum should have {\it systematically} fewer
OV-ionizing photons than predicted by \cite{haa96}, provided that
most HeII has been reionized to HeIII by the redshift in question.

\section{Metallicity of the Intermediate Density \lya Forest}\label{sec: civ}

\subsection{Identifying and Fitting CIV lines}

In \cite{hel97} we analyzed CIV artificial spectra in a manner similar
to \cite{son96}'s procedure, allowing a direct comparison with
\cite{son96}'s results.  However, the artificial spectra used were of
the ``segment spectra" variety, \ie, many small intervals of spectra at
particular redshifts, and the detailed modeling of Q1422 was not as
sophisticated as we do here.  In this section, we describe
a reanalysis of CIV absorption in Q1422, alongside an analysis of
CIV in our continuous artificial spectra.  The main additional
ingredient is that we now identify and fit CIV and HI absorption features in
the artificial spectra and Q1422 using the same routines.
Nevertheless, the basic results are similar to those of \cite{hel97}.

The automated Voigt profile fitter AutoVP (\cite{dav97a}) has been
extended to identify and fit metal line doublets
lying outside the \lya forest, such as CIV.  Once a spectrum has been
fitted with \lya lines, the identification scheme searches for spectral
features in the \lya forest region where the intensity drops below some
threshold, taken to be 0.7.  Within the wavelength interval where the
intensity is below that threshold, AutoVP searches for associated CIV
absorption in the two regions corresponding to the doublet positions.
If an absorption feature is due to CIV, it should be present in both
regions, with the strong component region (1548\AA) having twice the optical
depth of the weak component region (1551\AA) because of the difference 
in oscillator strengths.  
To identify any absorption that could possibly be CIV, AutoVP doubles the
optical depths of each pixel in the weak component region, superimposes them on
the strong line region with an appropriate shift in wavelength,
then takes the smaller of the two optical depths (strong component or
double the weak component) as its measure of CIV absorption.
Since CIV is the dominant 
species of absorption redwards of the \lya emission peak, the
implicit assumption that all absorption identified in this way is CIV
turns out to be quite good\footnote{It is even a reasonable
assumption for SiIV and NV, as the metal line region is typically
sparse enough that coincident absorption is rare.}.  We then fit this
remaining absorption with CIV Voigt profiles, and sum the column densities
within the region.  

The associated HI column density is found by summing all column
densities of HI lines within the associated spectral region.  Since the
\lya feature is typically saturated and therefore yields an inaccurate
estimate of the HI column density, AutoVP searches down the Lyman
series until it finds the first non-saturated Lyman series region, or
alternatively the last observable Lyman series region before the blue
cutoff in the spectrum; this is the HI line which is fit.  Wherever
possible, the unsaturated regions of other Lyman series lines (except
\lya) are converted to optical depths, and used to constrain the
optical depth at the associated position in the fitted Lyman series
line region; this minimizes the contribution due to coincident absorption
from other Lyman series lines.  After this constraint has been applied,
the Lyman series region is fit with Voigt profiles to obtain the HI
column density.  Often each region has several HI and CIV lines; we do
not attempt to associate individual CIV lines with HI lines, as varying
ionization conditions and bulk motions can make this 
identification very confusing and
produce unphysical results.  Our procedure is similar to that of
\cite{son96}, although they fit all Lyman series lines simultaneously
to obtain the HI column density.  In any case, we apply the exact same
detection and measurement algorithm to Q1422 and the artificial
spectra, so whatever biases are introduced will be similar in both data
sets.

\subsection{CIV/HI}

Figure~\ref{fig: CIVratio} shows our reanalysis of CIV lines in Q1422
and an identical analysis of our six artificial Q1422 spectra
assuming a uniform metallicity of [C/H]~$=-2.5$.  The identified CIV systems in
Q1422 correspond reasonably well with \cite{son96}'s systems; the main
differences arise from the estimate of $\nh$.  We conclude that our procedure
is not markedly different in practice from that of \cite{son96}.

The scatter in the CIV/HI ratio at a given column density 
in the simulated spectra arises from
spatial variations in the density and temperature and from variations
introduced by line identification and fitting. The ionizing background
and metallicity are assumed to be spatially uniform.  The assumed metallicity
of [C/H]~$=-2.5$ yields a mean CIV/HI ratio in good
agreement with Q1422, as we show in Figure~\ref{fig: metfrac}.  
Rauch et al. (1997a) and HDHWK found that a 1 dex scatter in
metallicity was required for simulations to reproduce the observed
scatter in CIV/HI.  Our more closely matched comparison here implies
a smaller intrinsic scatter of about 0.5 dex; the reduction arises
mainly because of additional ``observational'' scatter introduced
by the line fitting algorithm.

We determine the metallicity required to match the mean value of CIV/HI by
varying the assumed metallicity and then identifying and fitting metal lines in
the artificial spectra.  Figure~\ref{fig: metfrac}
(solid curve) shows the mean log column density of observed CIV lines
($\langle\log N_{\rm CIV}\rangle$) as a function of metallicity.  
We find $\langle\log N_{\rm CIV}\rangle$ to be the most robust statistic 
for this comparison.
Similar results are obtained by using the fraction
of HI lines with detectable CIV or the median value of CIV/HI.  The
solid horizontal line is $\langle\log N_{\rm CIV}\rangle$ 
for Q1422's CIV systems,
identified by the same method that was applied to the artificial spectra.  The
crossover is seen to occur right around [C/H]~=$-2.5$ (slightly higher
if the median CIV/HI is used).  The errorbars indicate the variance of
$\langle \log N_{\rm CIV}\rangle$ over the six artificial spectra; they do not
include the intrinsic scatter in CIV/HI.  The dotted curve will be
discussed in \S\ref{sec: hot} and the dashed curve in \S\ref{sec: cut}.

{}From this analysis, we confirm the results of 
Rauch et al. (1997a) and  \cite{hel97} that a mean
metallicity of [C/H]~$\approx -2.5$ with a small amount of spatial scatter,
together with an ionizing background of the \cite{haa96} shape,
reproduces the observed distribution of CIV/HI ratios in Q1422 quite
well.  Note that the cosmological model and $\Omega_b$ value used
for our simulations are substantially different from those used by 
\cite{hel97}, demonstrating the overall lack of sensitivity of
the mean metallicity determination to these assumptions.  We now turn
our attention to the question of whether this metallicity distribution
is consistent with observations of OVI absorption.

\section{Searching for OVI Absorption}\label{sec: search}

\subsection{Narrow Line Detection Algorithm}\label{sec: narrow}

Coincident HI absorption complicates the detection of
OVI(1032\AA,1037\AA) lines in quasar spectra.  For Q1422, OVI is
detectable only below about 4800\AA, lying in a region with a high
density of \lya(1216\AA) and \lyb(1026\AA) absorbers.

OVI lines are expected to be narrower than HI lines because thermal
broadening is a factor of four lower. 
However, as one goes to lower density structures,
broadening by residual Hubble flow across the absorbing region
becomes progressively more important (\cite{wei98}),
making OVI lines broader than naively predicted by thermal
broadening.  Nevertheless, we expect that the distribution of OVI line
widths will have a tail to smaller $b$-parameter (i.e., Gaussian velocity
width) than HI line widths.  Since our goal is to distinguish
OVI absorption from HI absorption rather than to identify all
possible OVI features, we focus on this tail of narrow OVI lines
in our detection algorithm.

Figure~\ref{fig: auto5} shows a comparison between a roughly
100\AA\ segment in one of our continuous artificial spectra versus the
spectrum of Q1422.  The top panel shows an artificial spectrum in the
same wavelength range without any metals added (i.e., only Lyman series
absorption).  The second panel shows the absorption caused purely
by OVI, including both doublet lines.  The third panel shows the
artificial spectrum with a metallicity of [C/H]$=-2.5$ and
[O/C]~$=+0.5$, essentially corresponding to the optical depth sum of
the first and second panels, since no other ions contribute
significantly in this wavelength range.  
The bottom panel shows Q1422 between 4510\AA\ and 4605\AA.
The noisy lines represent the data; the smooth curves will be explained
below.

{}From the third panel of Figure~\ref{fig: auto5} it is apparent that some
narrow features from OVI do survive despite the blanketing by HI.  We tried
a variety of algorithms to quantify the presence of these additional
narrow features.  The standard technique of decomposing the full
spectrum into a superposition of Voigt profiles (which are effectively
just Gaussian optical depth profiles in this low column density regime)
proved ineffective.  Many of the HI features are intrinsically
asymmetric, and in Voigt profile decomposition narrow components
are often introduced to fit the wings of these asymmetric lines.
It is therefore more useful to focus on {\it isolated} narrow lines.

As an alternative to Voigt profile decomposition, 
we developed a much simpler algorithm that searches
for spectral features that dip by more than $4\sigma$ and then return
to their original level all within a very small velocity interval, \eg
$18 \kms$ (4--5 pixels).  The $4\sigma$ criterion eliminates noise
spikes, while the velocity range criterion is tuned to optimally
select out the tail of narrow OVI lines that we would like to
identify.  We call this procedure our {\it narrow line detection algorithm}.
After identifying the narrow feature, we fit it using a Voigt
profile to obtain its column density
and $b$-parameter.
The identification and fits are shown as the smooth
curves in Figure~\ref{fig: auto5}.  As can be seen there, this algorithm
effectively picks out many of the narrow OVI features that
survive HI blanketing.

Occasionally, narrow lines are detected in the spectrum even without
any metals (top panel of Figure~\ref{fig: auto5}).  These narrow
lines arise primarily at velocity caustics, which
produce HI lines whose widths are close
to those expected from thermal broadening.  Figure~\ref{fig: auto5}
shows, however, that these false detections are significantly fewer in number
than the true OVI lines for 
an oxygen abundance of [O/H]$=-2.0$.  Figure~\ref{fig:
auto5} suggests that narrow lines are seen much
more rarely in Q1422 (bottom panel) than in an artificial spectrum with
$Z=-2.5$.  These qualitative impressions are borne out by the
statistics we present in the next section.

\subsection{Statistics of Narrow Lines}\label{sec: OVI}

We apply the narrow line detection algorithm to Q1422 between 
4025\AA\ and 4745\AA.  We apply the same algorithm over the
same interval to the six artificial Q1422 spectra,
first with no metals, then with a series of uniform metallicity
values ranging from $Z=-3.4$ to $Z=-2.2$.  At each metallicity we count the
number of narrow lines detected.  For Q1422, we detect 11 narrow lines,
while in the zero-metallicity case we detect an average of around seven
per artificial spectrum.  
Figure~\ref{fig: OVI} shows the number of narrow features detected
in the artificial spectra as a function of metallicity, quantitatively
confirming the inference from Figure~\ref{fig: auto5} that a 
metallicity of [C/H]$=-2.5$ with a relative overabundance [O/C]~$=+0.5$
is strongly inconsistent with the paucity of narrow features in the
spectrum of Q1422. At $Z=-2.5$, the expected number of narrow
line detections is $63\pm 8$ (where the $1\sigma$ error bars reflect
the variance over the six artificial spectra), which is inconsistent
with the Q1422 data at $\ga 6\sigma$ level.

We can strengthen this conclusion with an improvement to our
narrow line detection algorithm.  The narrow lines we identify as
candidate OVI lines should have a doublet companion if they are indeed OVI.  
While HI blanketing usually prevents actual identification of the
doublet companion feature, one can at least require that there be enough
absorption at the doublet position to obscure (or represent) the companion line.
Since we do not know which member of the doublet to assign to any
given narrow line, the stronger line at 1032\AA\ or the weaker line at
1037\AA, we must examine
both doublet wavelength positions; if neither contains
sufficient absorption, we can exclude that line from our list of OVI
candidates.

We can apply a similar technique to associated HI absorption.
{}From the LOX plot of Figure~\ref{fig: LOX}, one can see that OVI lines
are not expected to be detectable for systems with $\nh \la 10^{13}\cdunits$.
Below these densities, the amount of oxygen is small and much of it has
been ionized to OVII.  This $\nh$ cutoff is for $Z=-2.5$, and it scales
with metallicity.  Using this metallicity-dependent
relation, we can examine the spectrum at the two relevant positions to
see if there is sufficient absorption to allow a \lya line
of the appropriate HI column density.  If not, we can exclude that line.
The HI selection criterion removes more potential OVI detections than
the doublet companion selection criterion.

Applying these two additional constraints, we obtain the results plotted in
Figure~\ref{fig: OVIsel}.  The number of candidate OVI lines in Q1422
has dropped from 11 to 5 at $Z=-2.5$.  The number of lines excluded now
depends on the assumed metallicity, even for Q1422, since the
associated HI line strength required to exclude a narrow line as OVI
depends on the metallicity.  A slightly smaller number of lines ($\sim 4$
on average) are excluded from the no-metallicity artificial spectra
(when narrow lines are excluded by requiring $\nh > 10^{13}$, as at $Z=-2.5$),
making Q1422 compatible with having no narrow lines due to OVI.  A
slightly larger number of lines are excluded in the spectra with
metals.  At $Z=-2.5$, however, this larger number still represents 
a much smaller
fraction of the total number of narrow lines detected; the 
number of expected OVI detections is now $54\pm 6$, inconsistent
with Q1422 by $\ga 8\sigma$.
The number of narrow lines in the artificial spectra is
consistent with that in Q1422 only for metallicities $Z\la -3.2$.

In \S\ref{sec: O/C}, \S\ref{sec: hot}, and \S\ref{sec: cut} we examine
the sensitivity of this result to the various assumptions we have made
in our modeling.  Before that, we examine some
physical properties of the narrow lines identified by our algorithm.

\subsection{HI Column Densities Associated With Narrow Lines}\label{sec: Nhist}

We first focus on the artificial spectra with $Z=-2.5$.  
Even though this model is inconsistent with the Q1422 data,
it is helpful to examine the properties of its OVI absorbers
and associated HI and CIV lines in order to reinforce the points
made earlier in \S\ref{sec: OVIsearch}.

{}From the LOX plot of Figure~\ref{fig: LOX}, we claimed that OVI should
trace low column density (and hence low density) systems.  In order to
quantify this, we associate HI lines with each narrow feature in our
artificial spectra.  Since we cannot tell directly from an artificial
spectrum whether a feature is the stronger or weaker OVI component, we
sum the column densities of nearby HI lines within $25 \kms$ of the two
appropriate wavelengths, then use the greater of the two column
densities.  Owing to the high density of HI lines with $\nh\la
10^{14}\cdunits$, this is an inexact procedure, as it is
difficult to associate a single HI feature with a given OVI
line (even more so than in the case of CIV).  Nevertheless, summing all
the lines within $25 \kms$ (the typical width of HI features) provides a
reasonable estimate of the HI column density.  We also apply a similar
procedure to CIV absorbers identified in our artificial spectra.

We plot the histogram of associated HI column densities for OVI
(narrow) and CIV lines in Figure~\ref{fig: Nhist}.  OVI
lines trace HI with column densities down to $10^{13} \cdunits$ (associated
HI lines identified below this column density are probably spurious)
with a maximum sensitivity for systems around $10^{14} \cdunits$.
CIV traces systems with HI column densities down to $10^{14.5} \cdunits$,
with a maximum sensitivity around $10^{15} \cdunits$.  Thus, as we stated
earlier, OVI lines trace lower density regions, and the
``typical" HI column density traced by OVI absorbers is around
$10^{14} \cdunits$.  
The gas density in $\nh \approx 10^{14}\;\cdunits$ systems is close to the
mean cosmic baryon density.  Also as we stated earlier, the
majority of lines with detectable OVI absorption would have no
detectable associated CIV absorption.

\subsection{Column Densities and $b$-parameters of Narrow Lines}

In Figure~\ref{fig: bvsN} we plot the $b$-parameter of OVI-selected narrow 
lines (i.e., those tallied in Figure~\ref{fig: OVIsel}) against
the column density (inferred by assuming that they are HI lines).
In the upper left panel we show all the narrow
lines in the six artificial spectra with $Z=-2.5$.  In the lower left panel
we show the narrow lines identified in the artificial spectra with no
metals.  The lower right panel shows the ``extra" metal lines, i.e.
those in the $Z=-2.5$ spectra but not in the zero-metallicity artificial
spectra.  Finally, the upper right panel shows the OVI-selected narrow
lines in Q1422.

This figure shows that our narrow line detection algorithm effectively
picks out the tail of narrow lines that arise owing to the presence of OVI.  
The second panel of Figure~\ref{fig: auto5} (the artificial spectrum with
only OVI absorption) shows that many OVI features 
detectable by eye do not pass our narrow line selection criteria.
Still more OVI features are subsumed by HI absorption.  The OVI
lines that we do identify tend to be both strong and narrow.

There are many
more narrow OVI lines in the $Z=-2.5$ artificial spectra than seen in
the no-metallicity spectra.  If one further considers only lines with
$b< 10 \kms$ (dotted line in Figure~\ref{fig: bvsN}), there are
no lines in the zero metallicity case, while there are many
such lines in the $Z=-2.5$ case.  The simulation tests therefore
imply that any narrow line detected by this algorithm 
having $b<10 \kms$ has a good chance of being
a true OVI line, if not an intervening
metal line from a low redshift absorption system.

\subsection{Individual Narrow Line Systems in Q1422}

We now examine in detail the four narrow lines found in Q1422, shown in
the upper right panel of Figure~\ref{fig: bvsN}.  There is a fifth
narrow line, but it has $b\approx 25\kms$, and it appears to be due to a
single cold pixel in Q1422, so we discard it.  Two of the remaining lines have
$b<10 \kms$, making them strong candidates for OVI, while one line
(the circled point in Figure~\ref{fig: bvsN}) is the OVI system
identified by \cite{kir98}.  For each of these four systems, the doublet
position is subsumed by an HI line, so no direct confirmation of OVI 
absorption is possible.

Figures~\ref{fig: plotseg}(a-d) show the four narrow systems identified
in Q1422, along with the associated Ly$\alpha$, CIV1548, and CIV1551 spectral
regions.  For the various regions, the solid line corresponds to the
assumption that the narrow feature is OVI1032, while the dashed line
assumes OVI1037 absorption.  The other ions have been 
offset in flux for ease of viewing.

The first system, at 4276.5\AA, has a $b$-parameter of only $5.4
\kms$.  If it is OVI1032, it has an associated \lya line with
$\nh\approx 10^{14.5} \cdunits$ (confirmed by the \lyb component, not
shown), but no detectable CIV absorption (giving $N_{\rm CIV} <
10^{11.5}\cdunits$ at a $3\sigma$ level).  Since it is a very weak
line that is just above our $4\sigma$ detection threshold,
there is a possibility that it is an abnormally large noise
fluctuation.  It would be interesting to confirm this feature using
another spectrum of Q1422+231, \eg \cite{kir98}.  If it is an
OVI line, it has $N_{\rm OVI}\approx 10^{13.2}$, and we can determine 
the metallicity and oxygen overabundance
assuming an \cite{haa96} spectrum, the density-column density
relation from equation~(\ref{eqn: rho-NHI}), and the 
solar abundance values from Anders \& Grevesse (1989).
The ionization fractions are then 5.4\% for
CIV, 35\% for OVI, and $10^{-4.9}$ for HI, giving [O/H]~$=-2.5$ and
[C/H]~$\la -2.9$.  Thus this system, if OVI, would have a metallicity of
$Z\la -2.9$, with an overabundance of [O/C]~$\ga +0.4$.  These
values are quite consistent with the metallicity distribution inferred
from the statistics of narrow lines.

The second system (4278.9\AA) has $b=13.7 \kms$.  With only a modest
\lya line nearby (corresponding to this system being OVI1037), and no
CIV absorption, it seems quite likely that this system is a narrow HI line.

The third system (4517.2\AA) is a deep, narrow line with $b=7.5 \kms$.
However, there is a limited amount of corresponding \lya absorption, and no CIV
absorption.  For such a large OVI line ($N_{\rm OVI}\approx 10^{14}\cdunits$),
it is virtually inconceivable that there would be so little \lya and no
CIV absorption.  This contradiction favors an alternative explanation: 
this is an intervening
metal line from a low redshift absorption system.

The fourth and final system is the one presented in \cite{kir98} and
discussed in \S\ref{sec: OVIsearch}.  
The associated CIV absorption provides compelling evidence that this
is indeed an OVI line.  We measure a $b$-parameter of
$13\kms$, as opposed to $10\kms$ from \cite{kir98}.
Since we do not remove known \lyb absorption before
fitting, the \cite{kir98} value is probably more accurate for the OVI feature
itself.

All in all, of the four lines, only one is confirmed to be OVI (the
\cite{kir98} system), with another having some likelihood of being OVI.
This detailed analysis emphasizes the paucity
of detectable OVI absorption in Q1422.

\section{Varying the Theoretical Assumptions}\label{sec: assumptions}

\subsection{Overabundance and Uniformity of Oxygen}\label{sec: O/C}

Throughout this paper we have assumed an overabundance of [O/C]~$=+0.5$
compared with solar abundance ratios.  Had we assumed solar abundance ratios,
we would have favored the opposite conclusion, that a uniform metallicity
of [C/H]~$=-2.5$ is (within uncertainties) consistent with the OVI data.
This possibility therefore merits further 
discussion.

The theoretical motivation for oxygen overabundance is the assumption
that early star formation produced enrichment patterns characteristic
of Type II supernova yields, which have enhanced representation 
of $\alpha$-process elements.  High redshift enrichment almost 
certainly relies on Type II supernovae;
in a blowout model, both energetic and time
constraints require Type II supernovae (\cite{mir97}), while in a tidal
stripping model (\cite{gne97}; \cite{gne98}), there is insufficient 
time to produce a large number of Type I supernovae.  
Direct evidence for Type II enrichment patterns in the early
universe comes from local halo stars, which show an overabundance
of oxygen relative to carbon of 3--5 (\cite{edv93}), and
there is evidence that the very earliest stars are even more 
overabundant in $\alpha$-process elements (\cite{mcw97}).  
There is also mounting evidence for Type II enrichment patterns at high
redshift in forming galaxies (\cite{pet95}) and in damped \lya
absorbers (\cite{lu96a}).  Thus an overabundance of
[O/C]~$\sim +0.5$ for the IGM at redshifts $z\ga 3$ seems a 
reasonable, and perhaps even somewhat conservative, assumption.

We have also assumed a uniform metallicity.  From the CIV data, one sees
that at least some spatial scatter in metallicity is likely.
Such a scatter applied to the simulations would serve to {\it
strengthen} our conclusions, because it would tend to increase the
number of OVI systems in the strong-absorption tail of the distribution,
and these stronger lines are more likely to be detected despite coincident
HI absorption.  For spatial
variations in the ionizing background, the situation is more complex,
but a similar argument is still probably valid as long as there is no
systematic softening of the ionizing background above 114~eV.

\subsection{Reionization Heating}\label{sec: hot}

Our simulation assumed ionization equilibrium (balance between
recombination and ionization rates) at all times when computing
ionic abundances.  Departures from equilibrium abundances can,
under some circumstances, allow substantial heating of the IGM
during the reionization epoch (\cite{mir94}).
The low density gas that produces the \lya forest loses energy
mainly from adiabatic expansion, so the cooling timescale is comparable
to the Hubble time.  The relation between gas temperature and density
in the IGM therefore depends on the assumed reionization epoch and
associated heat injection (\cite{hui97}).
Our equilibrium abundance approach yields the temperature-density
relation that arises for minimal heating or a high reionization redshift.
The higher temperatures that arise in plausible alternative reionization
scenarios could in principle have a significant effect 
on our OVI modeling because the ionization fractions are sensitive to
temperature.

One example of the effect of different reionization assumptions
appears in figure~7 of Rauch et al. (1997b), which shows
the difference in the density-temperature relation between a
simulation run assuming ionization equilibrium (\cite{kat96}) and 
a simulation in which reionization occurred at $z\ga 6$ and the heat input was
tracked in a non-equilibrium manner (\cite{mir96}).  The latter
simulation shows hotter temperatures in the low-density regions, with
the discrepancy growing towards lower densities, as expected.

While we are currently unable to perform non-equilibrium simulations
of reionization, we can add the difference in the temperatures 
{\it post facto} to estimate the potential effect of the 
additional thermal energy.
Thermal pressure gradients are small compared to gravitational forces
in low density regions, so the dynamics of the gas should be unaffected
by higher temperatures, and the {\it post facto} change of the
temperature-density relation therefore accounts accurately for the full
effect of reionization heating.  In support of this claim,
we note that the distributions of the gas density for the two
simulations considered by Rauch et al.\ (1997b) are very similar (see their 
figure~6).

The difference in temperatures in the two Rauch et al.\ (1997b) simulations
can be approximated as
\begin{equation}
T_{\rm hot} = T_{\rm equil} 
2^{[1-\log(\rho/\bar\rho)]}\;\;\;\; {\rm for }\;\rho/\bar\rho < 10,
\end{equation}
where $\rho$ is the density and $\bar\rho$ is the mean baryonic density.
There is no difference for overdensities greater than ten, as the
cooling time there is sufficiently short to equilibrate the system.  By
adding this heat to each particle of our simulation before we construct
our artificial spectra, we obtain a set of ``reionization heated"
spectra.  For these spectra, the factor by which the \cite{haa96}
photoionizing background spectrum must be multiplied in order to
match the mean transmission of Q1422 is $0.86\pm 0.13$.

We perform the same CIV analysis on the reionization heated 
spectra to determine the mean metallicity that should be inferred from 
the observed CIV abundance in this hotter IGM model.
We find that reionization heating has a negligible effect (see
the dotted line in Figure~\ref{fig: metfrac})  because most of the
CIV absorption arises from systems with overdensities greater than
ten, translating to a column density $\nh \ga 10^{14.7}\cdunits$ (cf.
eqn.~[\ref{eqn: rho-NHI}]; also see Figure~\ref{fig: Nhist}).

For our OVI analysis, the results are somewhat more sensitive to 
reionization heating because OVI arises in lower density regions.  
If the IGM is hotter, more oxygen is ionized to OVII.  However, this is 
countered by the fact
that to match the mean transmission of Q1422 in the \lya
forest region, a hotter IGM requires a lower ionizing background
intensity.  A hotter IGM also means that the tail of OVI lines
below a given $b$-parameter is smaller, reducing the
efficiency of our narrow line search procedure for detecting OVI
and therefore increasing the OVI abundance inferred from a 
given number of detections.
This last effect turns out to be minimal; the
dominant effects are the first two, with the first being
somewhat more important.  
The results are shown in Figure~\ref{fig:
OVIhot} (solid line), indicating that the maximum metallicity allowed
for agreement with the OVI data is around $Z=-3.0$ instead of $Z=-3.2$.  
So, in
the final analysis, the effect of reionization heating does not
significantly change our overall conclusion that a uniform metallicity
of $Z=-2.5$ is inconsistent with the OVI data.

\subsection{Ionizing Background Spectrum}\label{sec: cut}

A more dramatic effect arises if the ionizing background contains many
fewer photons capable of ionizing oxygen to OVI.  Such a scenario might
arise if, for example, helium reionization has not occurred by 
$z\sim 3$.  We consider an idealized model of such a
scenario in this section.

To test the sensitivity of our analysis to the softness of the assumed ionizing
background, we generate a new $J_\nu$ with the shape from \cite{haa96},
except that the intensity is reduced by a factor of ten above 4~Ry; we
refer to this as the {\it cut} $J_\nu$.  We generate a new set of 
artificial spectra using this cut $J_\nu$; the mean transmission is
virtually identical to that of the original spectra because these high energy
photons do not significantly affect the HI opacity.
With the cut $J_\nu$, CIV is
much more common, and we find the mean metallicity for agreement
with CIV data to be [C/H]~$\approx -3.3$, as shown by the dashed line in
Figure~\ref{fig: metfrac}.  When we apply the same narrow line
detection procedure to spectra generated with the cut $J_\nu$, we obtain
the results presented in Figure~\ref{fig: OVIcut}.  This shows that a
metallicity of $Z<-2.9$ is in agreement with the OVI data.  
The cut $J_\nu$ scenario therefore implies
a mean metallicity in the \lya forest of around $Z\approx -3.3$.
With the assumed [O/C]~$= +0.5$, this scenario is consistent with
a uniform metallicity; however, we shall show in \S\ref{sec: pixmet}
that a cut $J_\nu$ also requires a much higher oxygen overabundance, 
thus a metallicity gradient is still required.

There is some evidence that helium reionization occurs around $z\approx 3$,
from a sudden drop in the SiIV/CIV ratio near that redshift
(\cite{son96}; \cite{son98}).  This evidence is controversial, as other
studies have not detected such a drop (\cite{bok98}),  
though \cite{son98} shows that it appears with several different
methods of analysis.
If many more
OVI lines were detected in quasar spectra at $z\la 3$, this would provide an
interesting argument in favor of the helium reionization scenario.
While Q1422 is poorly suited for this investigation, we hope that in
the future we can examine other quasar spectra over a wider range of redshifts.
Figure~\ref{fig: metfrac} implies that an increase in OVI due to 
a change in the ionizing background shape should be accompanied
by an increase in CIV/HI ratios.

One piece of evidence arguing that helium {\it has} been reionized by
this epoch is the measurement of the HeII opacity at $z\approx 3.3$
by \cite{hog97}.  From the strength of a HeII discontinuity about
5.3\AA\ from quasar Q0302-003, they find $\bar\tau_{\rm HeII}\approx
2^{+1}_{-0.5}$ ($2\sigma$ errors).  The \cite{haa96} spectrum we use
yields $\bar\tau_{\rm HeII}\approx 2.6$ at $z=3$, while the cut $J_\nu$
spectrum yields $\bar\tau_{\rm HeII}\approx 6.3$, well outside
the quoted uncertainty of the observed optical depth.
However, the observation of patchy HeII absorption at $z \sim 3$
by \cite{rei97} suggests that HeII reionization
does occur near this redshift, and that the \cite{hog97}
measurement in the immediate vicinity of Q0302 might
provide a poor estimate of the overall mean HeII absorption.
A more secure determination of $\bar\tau_{\rm
HeII}$ at $z\ga 3$ may have to await the deployment of the
Cosmic Origins Spectrograph on the Hubble Space Telescope.

\section{Detection of OVI by the Optical Depth Ratio Technique}\label{sec: pixmet}

\subsection{Algorithm and Q1422 Results}\label{sec: pixel}

Our narrow line detection results are consistent with either
the metallicity being lower in lower density regions or the ionizing
background being significantly truncated above 4~Ry (assuming a
reasonable overabundance of oxygen relative to carbon).  In order to
distinguish between these two scenarios, we apply a different algorithm for
detecting OVI that was developed by \cite{son98}.  In this
algorithm, no metal line identification is done; rather, the presence
of metals is assessed statistically by a pixel-by-pixel search for
excess optical depth at the expected position of associated absorption.  
In this section we describe this {\it optical depth ratio} technique.

Using this technique, we search for OVI absorption
associated with all regions of space showing significant CIV
absorption.  For all pixels redwards of the quasar's \lya emission
peak, we first identify pixels that have $\tau > 0.05$.  From these we
select pixels consistent with absorption arising from the weak doublet component
of CIV (1551\AA).  Specifically, we only accept a pixel if the optical
depth at the strong doublet component's position (1548\AA) is at least
double the optical depth at the weak component's position (with some
allowance for noise).  The final subset of pixels represents all
regions along the line of sight to the quasar showing significant CIV
absorption.  For Q1422, 376 out of a total of roughly 25,000 pixels
redwards of the \lya emission peak are selected as having significant
CIV absorption.  For each of the selected pixels, we find the optical
depth at the associated OVI position and calculate the
``apparent column density ratio":
\begin{equation}
{N_{\rm OVI,app}\over N_{\rm CIV}} = 
\left({\tau_{\rm OVI} \over \tau_{\rm CIV}}\right)
\left({\lambda_{\rm CIV}\; f_{\rm CIV}\over
\lambda_{\rm OVI}\; f_{\rm OVI}}\right),
\end{equation}
where $\tau$ is the optical depth, $\lambda$ is the
rest wavelength, and $f$ is the oscillator strength for each ion.
$N_{\rm OVI,app}$ is not the true OVI column density because there
is a significant contribution from coincident HI absorption.

Although the LOX plot of Figure~\ref{fig: LOX} shows that the
detectability of CIV and OVI peaks at different HI column densities,
it also shows that the weaker (and therefore more common) CIV lines,
associated with $\nh\sim 10^{15}\cdunits$,
should have significant amounts of photoionized OVI.  This is seen when
we apply the optical depth ratio technique to artificial spectra with
oxygen ([O/H]~$=-2.0$) and without oxygen, the latter being used to
assess the contribution from coincident HI absorption.  The results are
shown in Figure~\ref{fig: pixmet}, as a histogram of $N_{\rm
OVI,app}/N_{\rm CIV}$ (this is similar to Figures~20e-g from
\cite{son98}, except that \cite{son98} uses randomly selected regions
of the observed spectrum instead of a zero-metallicity model spectrum
as the ``control'' sample).  There is a significant excess of OVI seen in the
[O/H]~$=-2.0$ spectra (solid line) as compared to the spectra without
no oxygen (dotted line); this is mostly due to photoionized OVI
associated with CIV absorbers, as the covering fraction of
collisionally ionized OVI is quite small.  We also apply the technique
to Q1422, and find a significant presence of OVI, consistent with
\cite{son98}.  The Q1422 results (dashed histogram in Figure~\ref{fig:
pixmet}) are seen to be in better agreement with the $Z=-2.5$ spectra than
with the zero-metallicity spectra.
We will quantify these results further in the next section.

Our conclusion from optical depth ratios that [O/H]~$\approx -2.0$
may seem inconsistent with our earlier conclusion from narrow lines
that [O/H]~$\leq -2.5$, but it is not, {\it provided} that the
IGM metallicity is lower at lower gas densities.
With the narrow line detection algorithm, we are examining {\it all}
regions of space for OVI absorption, so the measure is weighted
towards the low-density regions that occupy most of the volume. Conversely,
the optical depth ratio technique examines regions of
space selected to have significant CIV absorption, which are typically
higher density regions.  In fact, from \S\ref{sec: civ} we {\it know}
the metallicity in these regions is $Z\approx -2.5$ in the full HM
spectrum case, or $Z\approx -3.3$ in the case of our cut $J_\nu$ spectrum
(cf. \S\ref{sec: cut}).  Thus, we can use the detection of OVI in these
regions as a probe of the shape of the ionizing background and the
assumed oxygen overabundance, as we will show in the next section.

For regions that have detectable CIV absorption,
the optical depth ratio technique is significantly more sensitive to
the presence of OVI than the narrow line detection algorithm.  
Because 
most OVI positions are
contaminated by HI absorption, the opportunity to detect a narrow
OVI line associated with CIV is very rare.  
The optical depth ratio technique can still achieve a statistical
detection of OVI in such cases because it examines only the precise
locations of expected OVI absorption and compares their mean optical
depth to an OVI-free control sample.

We attempted to quantify the presence of OVI associated with \lya
absorption of a given optical depth, but were unsuccessful.
The optical depth ratio technique is much less reliable when used with
HI rather than CIV, since the typical line widths of HI features are
significantly different from those of CIV or OVI; it is only because
OVI and CIV have similar thermal broadening (and identical bulk flow
broadening because we are examing the same physical region) that
optical depths at associated wavelengths can be meaningfully compared.

\subsection{Constraining Ionization Conditions and Metallicity Gradient}

We now apply the optical depth ratio technique to artificial spectra
with varying metallicity.  For the artificial spectra, the assumed
[C/H] is held fixed at the value that produces agreement with the CIV
data (\ie $-2.5$ for the \cite{haa96} spectrum and $-3.3$ for the cut
spectrum), and we vary only [O/H] (or equivalently, the oxygen
overabundance).  We plot the median value of $N_{\rm OVI,app}/N_{\rm
CIV}$, since the median provides the most robust characteristic
quantity in this highly skewed distribution.  The median $N_{\rm
OVI,app}/N_{\rm CIV}\approx 8.3$ for Q1422, and $\approx 2.6$ for the
artificial spectra without oxygen.  These values are shown as the
horizontal dashed and dotted lines in Figure~\ref{fig: Zpix},
respectively.  The varying [O/H] case is shown by the solid line, with
the error bars as usual computed over the six artificial spectra (the
error bars on the other artificial spectra measurements are
comparable).  Agreement with Q1422 is seen to occur at [O/H]~$\approx
-2.0$, and given that the metallicity of these regions is
[C/H]~$\approx -2.5$, this implies [O/C]~$\approx +0.5$.  Thus for the
case of the HM spectrum, we have independently determined that our
assumed oxygen overabundance is consistent with Q1422.

We now examine the cut $J_\nu$ spectrum case, as shown by the
dot-dashed line in Figure~\ref{fig: Zpix}.  With a softer spectrum, OVI
is much less abundant and CIV is more abundant, meaning that a significantly
higher overabundance of oxygen is required to reach agreement with the
Q1422 data for OVI associated with CIV regions.  Agreement is reached
with Q1422's median $N_{\rm OVI,app}/N_{\rm CIV}$ for [O/H]~$\approx -1.0$,
which when combined with [C/H]~$\approx -3.3$ for the cut spectrum
implies [O/C]~$\approx +2.3$.  A factor of $\ga 100$ oxygen
overabundance seems fairly implausible, as no set of objects has ever
been observed to have $\alpha$-process overabundances
approaching this level.

We conclude that the regions showing CIV absorption are in much better
agreement with a \cite{haa96} spectrum with no softening above 4~Ry,
and therefore that most helium is doubly ionized by redshift $z\sim 3.6$.
This scenario is also consistent with an oxygen overabundance of
[O/C]~$\approx +0.5$.

free We can assess the presence of a gradient of metallicity with
density, free from assumptions about the oxygen overabundance, by
comparing [O/H] 
from the narrow associated with CIV regions with the limits on [O/H] in
lower density gas from the narrow line detection algorithm.  For the
\cite{haa96} spectrum case, CIV regions have [O/H]~$\approx -2.0$,
while the narrow line search 
[O/H]~$\la -2.5$, implying at least a factor of 3 decrease in
metallicity at lower densities.  For the cut $J_\nu$ spectrum, CIV
regions have [O/H]~$\approx -1.0$, whereas the narrow line search
yields [O/H]~$\la -2.9$, implying an even {\it greater} metallicity
gradient.  Thus regardless of the assumed ionizing background, a
metallicity gradient is required to match the observed amount of OVI
from both detection methods, barring the unlikely scenarios that the
metagalactic ionizing flux is significantly {\it harder} than the
\cite{haa96} spectrum, or the oxygen overabundance is significantly
lower in the lower density regions.

\subsection{IGM with Patchy Ionization Conditions}

As mentioned before, \cite{son98} detected a significant jump in 
SiIV/CIV ratio above $z\approx 3$, that when combined with the SiIV/CIV
ratio above $z\approx 3$, which, when combined with photoionization
models, is consistent with an ionizing spectrum heavily truncated (by a
factor $\ga 100$) above 4~Ry.  This inference is in direct
contradiction with the results of the previous section.  One way to
reconcile these results may be to invoke an IGM that has a patchy
ionization structure.  This type of structure is implied by the HeII
measurements of \cite{rei97}, as stated in \S\ref{sec: cut}, and has
been suggested by \cite{son98} to explain the significant presence of OVI
detected at $z\ga 3$.  In this section we consider a simplistic,
``50/50'' model of such a patchy IGM to determine the implications for
our OVI detections.

Consider a universe where half the volume has already reionized helium
(and hence has an \cite{haa96} spectrum), and half the volume has a
significantly truncated ionizing flux above 4~Ry (as in our cut $J_\nu$
spectrum).  
CIV absorption then implies a mean metallicity of [C/H]~$\approx -2.9$.  
This value can be determined from Figure~\ref{fig: metfrac}, as
this is the metallicity where the solid (\cite{haa96} spectrum) and
dashed (cut spectrum) curves are roughly equally above and below the
observed $<\log(N_{\rm CIV})>$ of Q1422.  The additional scatter
introduced by the varying ionization conditions implies that there
can be virtually no intrinsic spatial scatter in the metallicity of
CIV absorbers.
Similarly, bisecting the solid and dot-dashed curves in Figure~\ref{fig: Zpix}
implies agreement with the Q1422 optical depth ratio results at
[O/H]$\approx -1.7$.  
Taken together, the [C/H] and [O/H] constraints imply [O/C]~$\approx +1.2$
in the 50/50 model.
This overabundance is perhaps marginally plausible if
one invokes enrichment almost entirely from high-mass stars yielding
a high fraction of $\alpha$-process elements.  

The narrow line detection results become statistically consistent with
the Q1422 data only for [O/H]~$\la -2.5$ (metallicity $Z \la -3$),
as seen by bisecting the solid and dot-dashed curves in 
Figure~\ref{fig: OVIcut}.
Thus, a metallicity gradient is still required in the 50/50 model,
as the optical depth ratio technique
(tracing CIV regions only) yields [O/H] higher by $\approx +0.8$~dex than
the narrow line detection technique (tracing mostly lower density regions).

While the 50/50 patchy IGM model is marginally consistent with the OVI
data, it remains to be seen whether such a scenario is consistent with
the observations of SiIV/CIV by \cite{son98}.  Currently, with only one
quasar spectrum available to us, we do not have sufficient statistics
to investigate SiIV in detail; we plan to conduct these investigations
when more data become available, preferably at $z\la 3$ as well as at
$z\ga 3$.

\section{Discussion and Conclusions}\label{sec: disc}

We present a systematic search for OVI absorption in the spectrum of
Q1422+231 ($z=3.62$), using a narrow line detection algorithm proven
effective at identifying OVI absorption in artificial spectra, and an
optical depth ratio technique introduced by \cite{son98}.  The first
technique traces OVI predominantly in systems with $10^{13.5}\la \nh\la
10^{15}\cdunits$, whereas the second technique traces only OVI 
associated with CIV absorption, \ie in $10^{14.5}\la \nh\la
10^{16}\cdunits$ systems.  By comparing Q1422 and artificial spectra 
having varying metallicities, we determine that
\begin{enumerate}

\item{[O/H] must be lower in lower density regions, for 
either an
\cite{haa96} ionizing background or an ionizing background
significantly truncated above 4~Ry.  If [O/C] is constant in systems up
to $\nh\sim 10^{16}\cdunits$, our results imply that regions traced by
$\nh\la 10^{14}\cdunits$ systems (corresponding to gas at roughly the
mean baryonic density) have a mean metallicity lower by at least a
factor of 3 compared to regions traced by $\nh\sim 10^{15}\cdunits$
systems (corresponding to a baryonic overdensity of $\sim 10$).}

\item{More than half the universe must have helium reionized by $z\sim 3$.  
If helium has completely reionized by $z\sim 3.6$ (the highest redshift probed
by the Q1422 data), then
our analysis implies [O/C]~$\approx +0.5$, in good agreement with
overabundance measurements of Type II supernovae enriched systems.  If
a significant portion of the universe has not reionized helium by
$z\sim 3$ and therefore has a softer ionizing background spectrum, 
then the required oxygen overabundance is higher.
For example, if half of the volume has not reionized helium, then
[O/C]~$\approx +1.2$, already greater
than the observed overabundance of any class of Type II supernovae
enriched systems.
If the spectrum were soft throughout the universe at $z\ga 3$ then an
implausibly high overabundance, [O/C]~$\approx +2.3$, would be required.
}
\end{enumerate}

These conclusions are in good agreement with the recent study of
\cite{lu98}, who used composite spectra to investigate CIV absorption
in systems with $10^{13.5}< \nh < 10^{14} \cdunits$ and found that the
metallicity of these absorbers must be [C/H]~$\la -3.5$.
Cosmological simulations show that $\nh \sim 10^{14} \cdunits$ roughly
corresponds to the dividing line between overdense and underdense
regions of the universe (though the value of $\nh$ that marks this
division depends on redshift and, to a lesser extent, on cosmological
parameters).  Our results therefore imply that mildly overdense
regions such as filaments and sheets have been enriched,
while underdense regions are virtually chemically pristine.  Simulations
that self-consistently enrich the IGM by tracking metal production and
transport find that a strong metallicity gradient is predicted between
the mildly overdense and underdense regions (see figure~3 in
\cite{gne98}); this predicted gradient is in good agreement with the
\cite{lu98} data and with the scenario we present above.

Recent measurements of a jump in the SiIV/CIV ratio around $z\sim 3$
(\cite{son98}) may be difficult to reconcile with conclusion~(2) above.  While 
we have yet to conduct a systematic comparison of SiIV in observed and
artificial spectra, primarily because of the small numbers of SiIV systems
detectable in our one available quasar spectrum, we expect our results
will be in agreement with \cite{son98}, who argues that such a jump
requires a much softer ionizing background at $z\ga 3$.  One way to
reconcile these results may be to invoke patchy helium reionization at
that epoch, as suggested by \cite{rei97}; such a model may be tested in
greater detail by searching for an anti-correlation between OVI and
SiIV detections.

If helium reionization occurs around $z\sim 3$, our narrow line
algorithm should yield many more OVI detections at redshifts $z\la
3$.  Such searches are difficult because of the poor blue sensitivity
of the HIRES spectrograph (and complete loss of sensitivity at
$\lambda\la 3800$\AA), but quasar spectra do exist that could provide
constraints down to $z\sim 2.7$.  The presence of a substantial number
of OVI lines in this regime would strongly favor the late helium
reionization scenario; there is already some weak evidence that OVI is
more abundant at $z\la 3$ (\cite{son98}).  If OVI features continue to
be virtually undetectable down to $z\sim 2.7$, this would be compelling
evidence against the late helium reionization scenario, since the HeII
absorption measurements of Davidsen, Kriss, \& Zheng (1996) imply that
helium has been reionized by this redshift.  We hope to work with
observers to attempt this search in the near future.

In a broader context, our work illustrates the power of combining
cosmological hydrodynamic simulations of structure formation with
high-quality quasar spectra to infer the ionization state and
the enrichment history of the high-redshift IGM.  Future observations
and simulations promise a wealth of information, which, when combined,
will help us to better understand the evolution of the IGM and its
connection to early star formation and the epoch of primeval galaxies.

\acknowledgments
We are very grateful to Antoinette Songaila and Len Cowie for making
their Keck quasar spectrum of Q1422+231 available to us.  We thank Eric
Linder and Antoinette Songaila for helpful discussions.  We also thank
Francesco Haardt and Piero Madau for providing us with their latest
ionizing background in electronic form.  UH acknowledges support from a
postdoctoral research grant from the Danish Natural Sciences Research
Council.  This work was supported in part by the PSC, NCSA and SDSC
supercomputing centers, by NASA theory grants NAGW-2422, NAGW-2523,
NAG5-2882, NAG5-3111, and NAG5-3820, by NASA LTSA grant NAG5-3525, by
NASA HPCC/ESS grant NAG 5-2213, and by the NSF under grants ATS90-18256
and ASC 93-18185.  Finally, we acknowledge the essential contribution
of S. Vogt and the HIRES team in building the instrument that made this
kind of investigation possible.


\clearpage


%
%

\clearpage

\figcaption[fig.spec.ps]{ {\it Top left panel:} Q1422, with continuum fit 
and $4\sigma$ noise level shown for the \lya forest region.
{\it Bottom left panel:} One of the six continuous artificial spectra,
with continuum fit and $4\sigma$ noise level shown. {\it Top right panel:} 
\lya forest region of Q1422 after continuum normalization, with $4\sigma$
noise level.  {\it Bottom right panel:} Normalized \lya forest region of 
the artificial spectrum, with $4\sigma$ noise level.
\label{fig: qspec}}

\figcaption[fig.LOX.ps]{Line Observability Index (LOX; see
\cite{hel98}) for OVI and CIV, with [C/H]~$=-2.5$ and [O/C]~$=+0.5$.  Solid curves are obtained by using
the \cite{haa96} spectrum; dashed curves use the ``cut" spectrum
described in \S\ref{sec: cut}.  The horizontal dotted line represents
(roughly) the detectability limit redwards of the quasar \lya peak,
while the horizontal dot-dashed line represents the detectability bluewards.
Note the increase in CIV detectability
and decrease in OVI detectability using the cut spectrum.
\label{fig: LOX}}

\figcaption[fig.CIVratio.ps]{CIV/HI ratios for all lines with $\nh >
2\times 10^{14} \cdunits$, in Q1422 (solid squares) and six continuous
artificial spectra (open circles).  The identical CIV identification
and Voigt profile fitting routine has been applied to both data sets.
\label{fig: CIVratio}}

\figcaption[fig.metfrac.ps]{The variation of $<\log N_{\rm CIV}>$ with
metallicity for the artificial spectra, as compared with the value obtained
for Q1422 (horizontal line).  For the spectra using the \cite{haa96}
background, the agreement is achieved for $Z\sim -2.5$.  The result
is virtually unchanged by adding heat due to reionization (see \S\ref{sec: hot}),
but is dramatically different ($Z\sim -3.3$) using the ``cut" 
spectrum described in \S\ref{sec: cut}.
\label{fig: metfrac}}

\figcaption[fig.auto5.ps]{ 4510\AA\ to 4605\AA\  segment from Q1422 and from one of the
artificial spectra.  Narrow lines have been identified and fitted (shown as smooth curves) according
to the prescription described in \S\ref{sec: narrow}.
{\it Top panel:} Zero metallicity.  {\it Second panel:} Only OVI
absorption, with [O/H]~$=-2$ (both doublet components).  {\it Third panel:}
Artificial spectrum with $Z=-2.5$.  {\it Bottom panel:} Q1422.  Note
the observability of narrow lines in the artificial spectrum, as
compared with Q1422.
\label{fig: auto5}}

\figcaption[fig.OVI.ps]{The number of narrow lines detected in the
artificial spectra with varying metallicity, versus the value for Q1422
(11 lines) and the artificial spectra without metals (7 lines).
Agreement with Q1422 is reached only for metallicities around
$Z\approx -3.0$.
\label{fig: OVI}}

\figcaption[fig.OVIsel.ps]{The number of narrow lines selected for OVI
compatibility using the doublet and HI criteria described in \S\ref{sec: OVI}.
At $Z=-2.5$, Q1422 has five possible OVI lines, while the no-metallicity
artificial spectra have on average four.  Consistency with Q1422 is reached
only for metallicities $Z\la -3.2$.  This metallicity implicitly assumes
an overabundance of oxygen [O/C]~$=+0.5$, as expected for Type II
supernovae-enriched gas.
\label{fig: OVIsel}}

\figcaption[fig.Nhist.ps]{Histogram of column densities of associated
HI lines for OVI (narrow) lines and CIV lines, from the artificial
spectra with $Z=-2.5$.  The OVI lines trace lower density gas, with
a peak sensitivity around $\nh \approx 10^{14}\cdunits$, while 
CIV lines have a peak sensitivity around $\nh \approx 10^{15.5}\cdunits$.
In this scenario, most OVI lines detected will have no associated CIV
absorption.
\label{fig: Nhist}}

\figcaption[fig.bvsN.ps]{Scatter plot of $b$-parameters vs. column
densities (fit assuming narrow line is HI) for the narrow lines 
identified in the six $Z=-2.5$ artificial spectra and selected for OVI, 
as compared with Q1422.  {\it Upper left:} All narrow lines.
{\it Lower left:} Narrow lines in zero-metallicity spectra.
{\it Lower right:} Extra narrow lines due to the presence of OVI.
{\it Upper right:} Narrow lines in Q1422.  Note that no narrow
lines with $b<10 \kms$ appear in the spectra without metals.
\label{fig: bvsN}}

\figcaption[fig.plotseg1.ps]{(a--d) Four individual candidate OVI
systems in Q1422, with associated \lya and CIV(1548\AA,1551\AA)
regions.  The solid line shows the region if the narrow line is
OVI(1032\AA), while the dashed line corresponds to assuming the narrow
line is OVI(1037\AA).
\label{fig: plotseg}}

\figcaption[fig.OVIhot.ps]{The number of narrow lines selected for OVI in
artificial spectra with reionization heat added as described in 
\S\ref{sec: hot}.  Agreement with Q1422 occurs at slightly higher metallicity,
$Z\la -3$.  This is still inconsistent with a uniform metallicity
of $Z=-2.5$ (assuming an overabundance of oxygen [O/C]~$=+0.5$).
\label{fig: OVIhot}}

\figcaption[fig.OVIcut.ps]{The number of narrow lines selected for OVI
in artificial spectra with a spectrum cut by a factor of ten above
4~Ry, as described in \S\ref{sec: cut}.  Agreement with Q1422 occurs
for $Z\la -2.9$.
\label{fig: OVIcut}}

\figcaption[fig.pixmet.ps]{Histogram of $N_{\rm OVI,app}/N_{\rm CIV}$
in pixels associated with significant CIV absorption, as identified and
measured by the optical depth ratio technique (\S\ref{sec: pixel}).
Solid line shows the results from six artificial spectra with
[O/H]~$=-2.0$ and [C/H]~$=-2.5$, the dotted line shows the results from
artificial spectra without oxygen, and the dashed histogram shows the
results from Q1422.
\label{fig: pixmet}}

\figcaption[fig.Zpix.ps]{Median $N_{\rm OVI,app}/N_{\rm CIV}$ versus
[O/H], with [C/H] held fixed at the value producing agreement with CIV
data (solid line).  The dashed line shows the Q1422 value of 8.3 while
the dotted line shows the value from artificial spectra without oxygen
(2.6).  This confirms \cite{son98}'s observation of a significant
presence of OVI.  The dot-dashed line shows the results from artificial
spectra generated using the cut spectrum described in \S\ref{sec:
cut}.
\label{fig: Zpix}}



\clearpage
\plotone{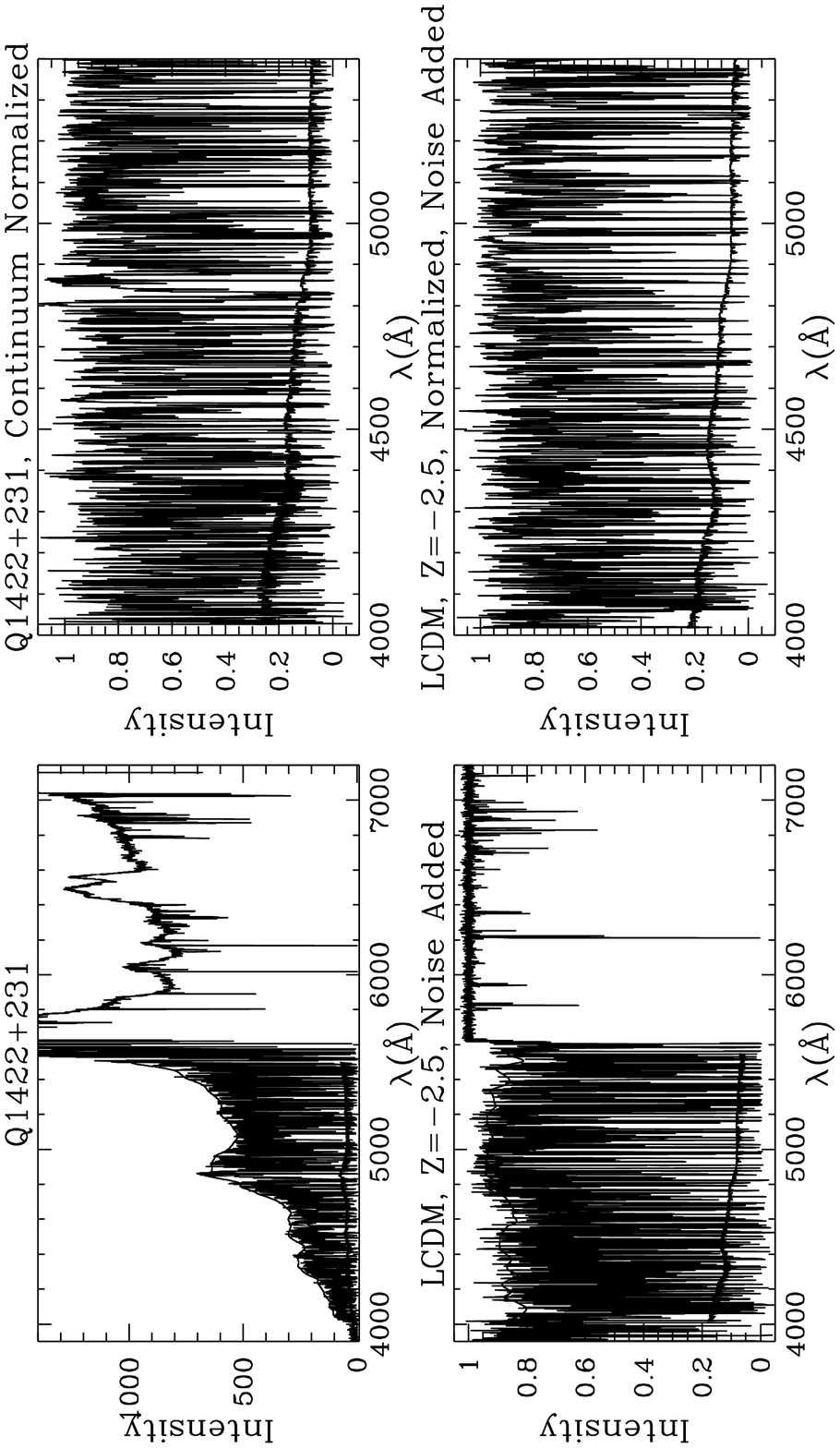}

\clearpage
\plotone{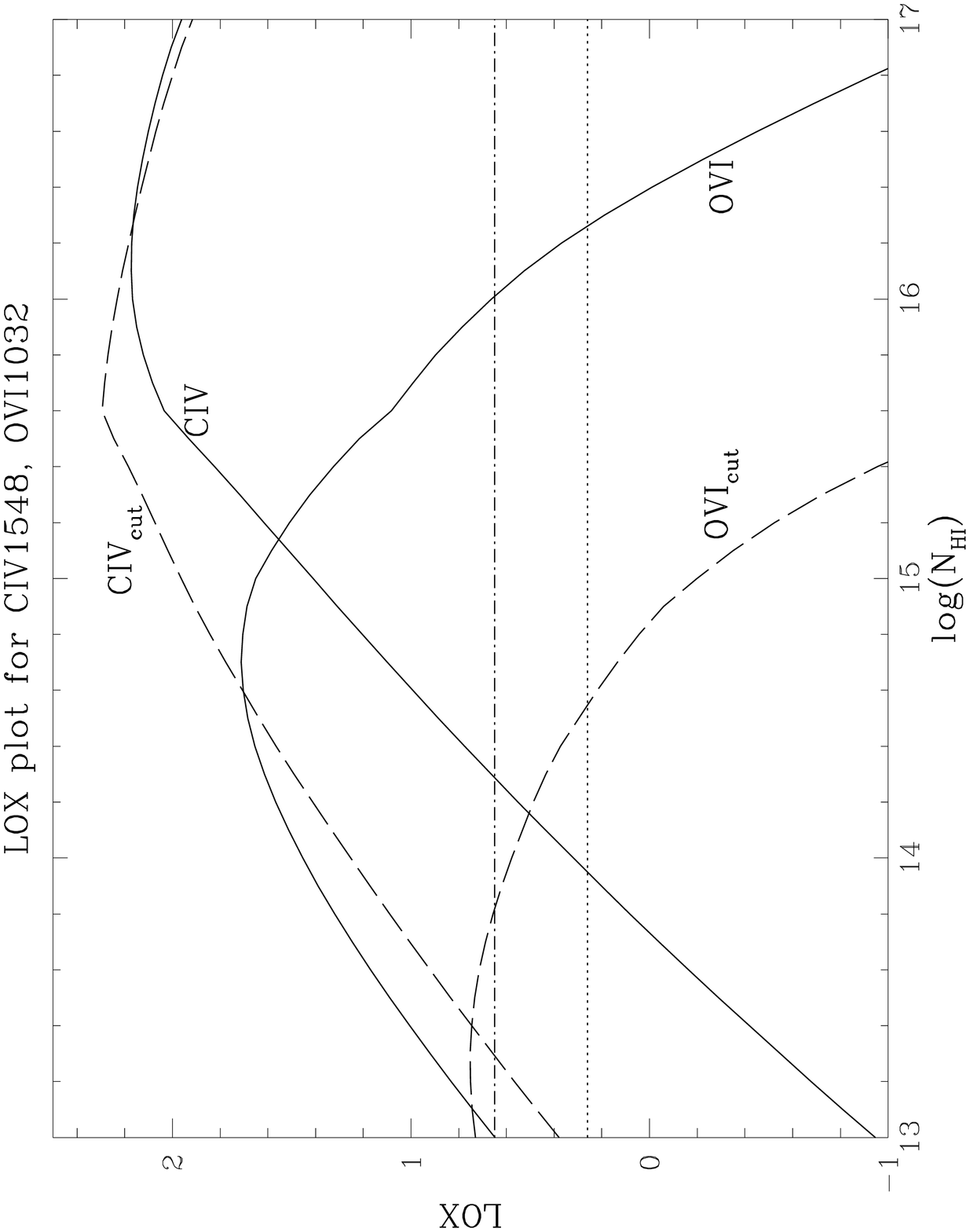}

\clearpage
\plotone{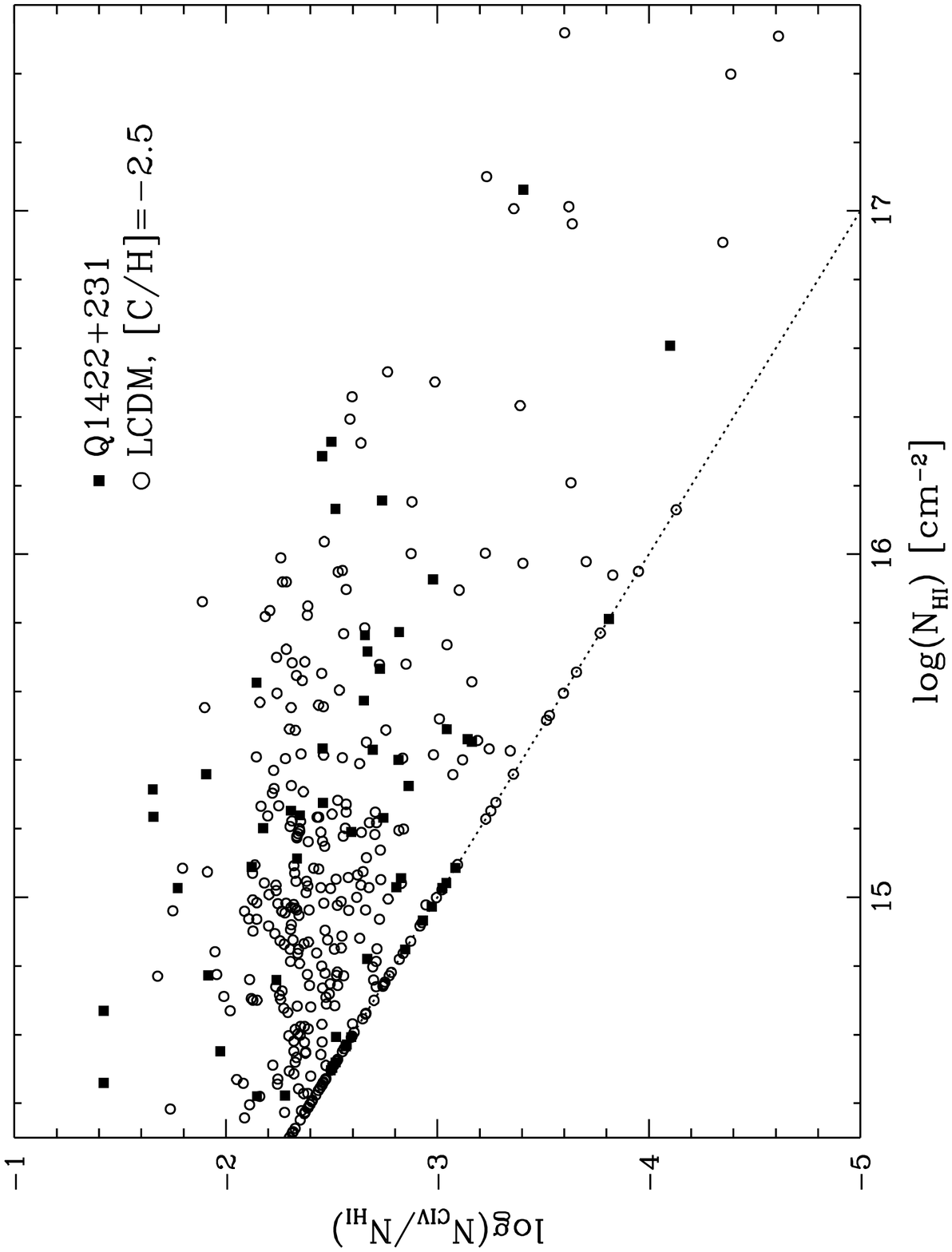}

\clearpage
\plotone{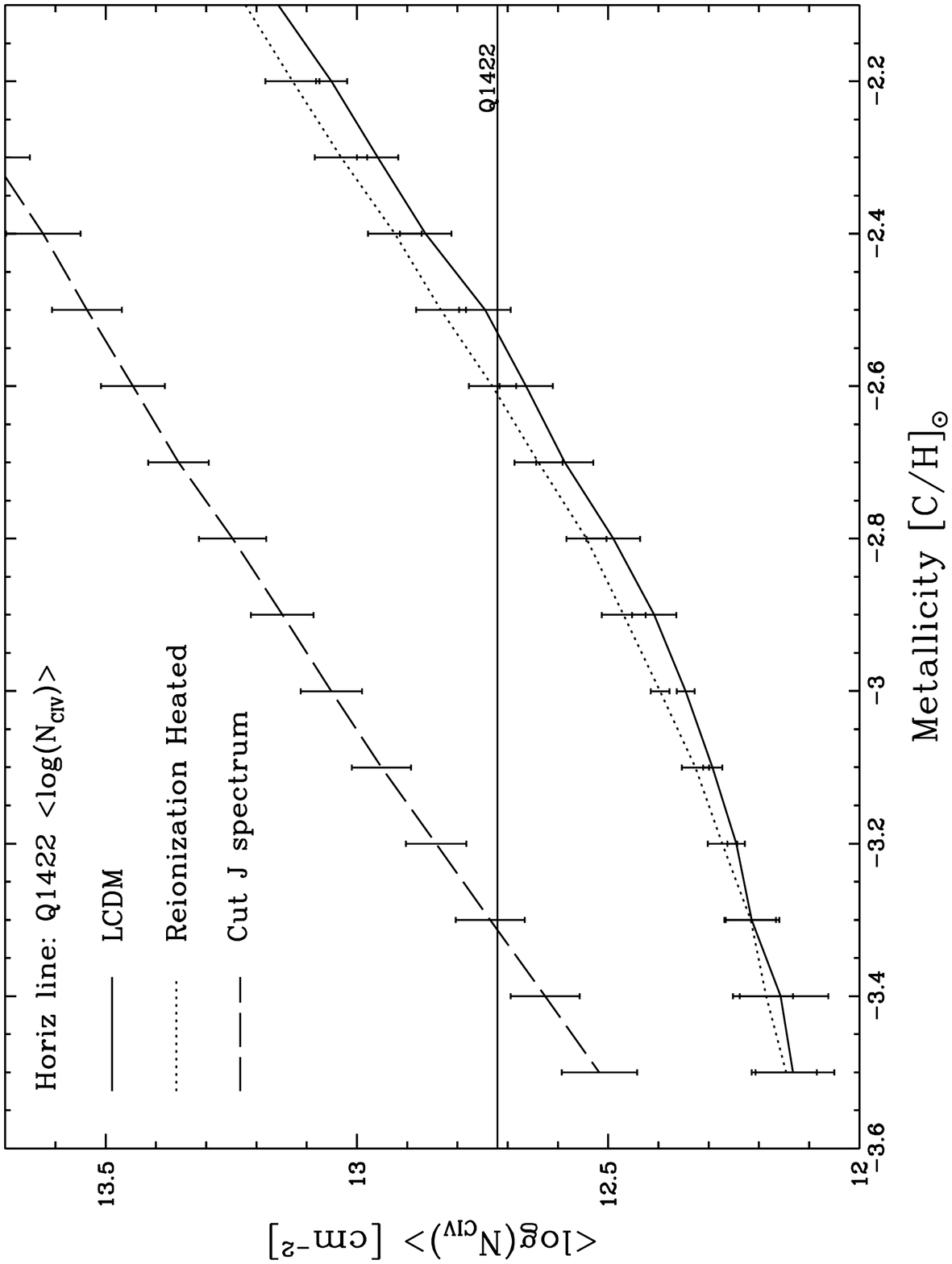}

\clearpage
\plotone{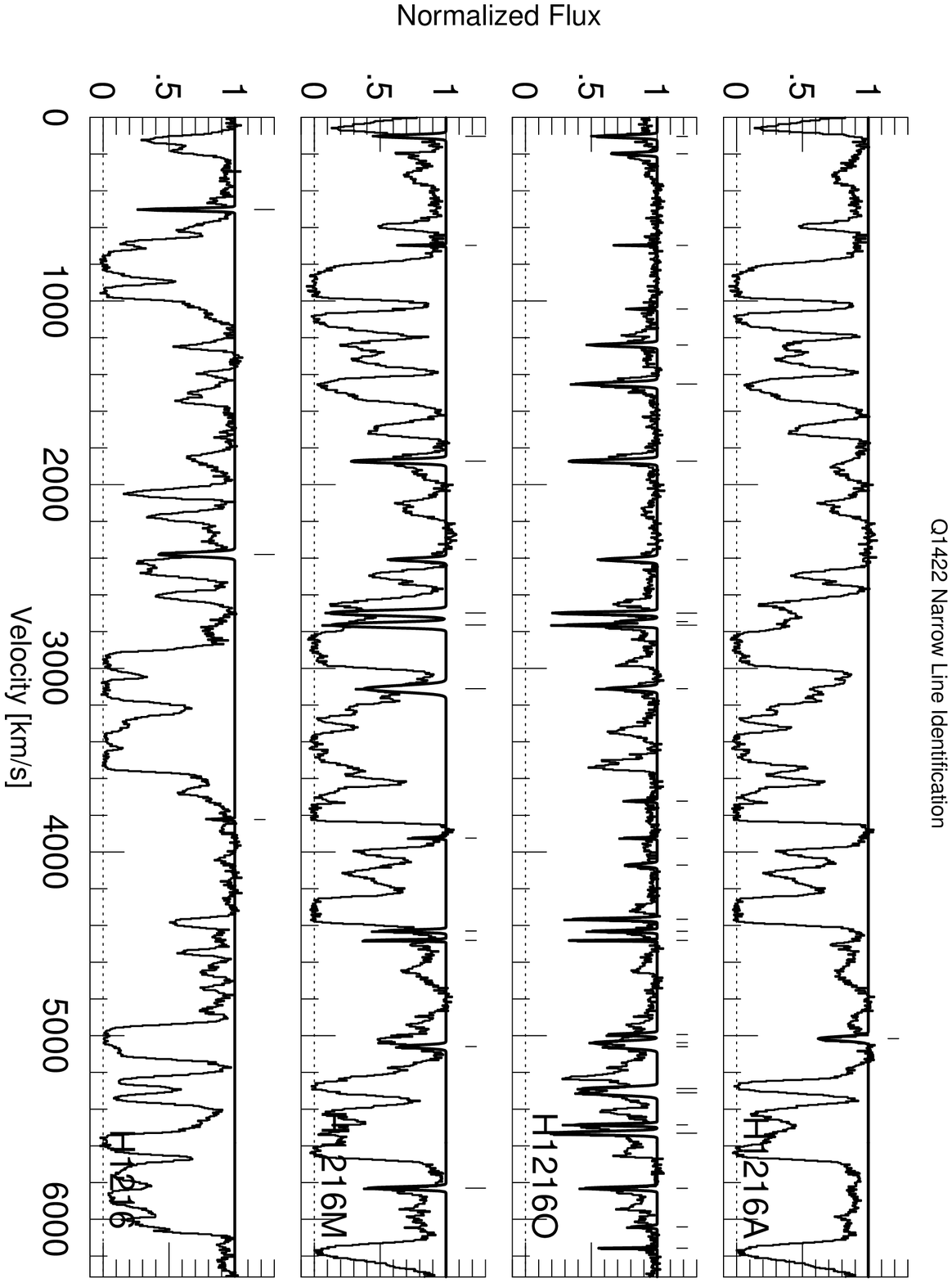}

\clearpage
\plotone{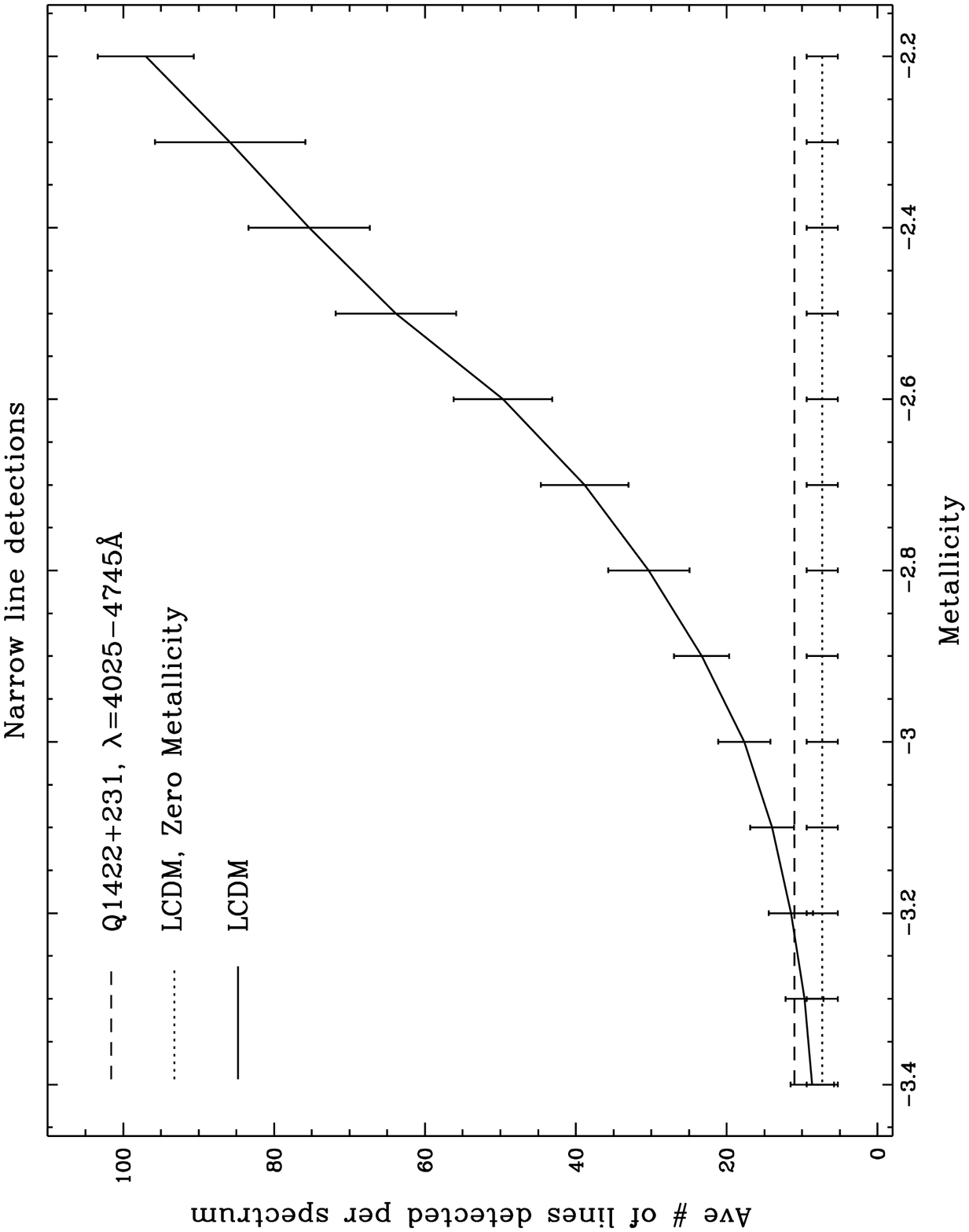}

\clearpage
\plotone{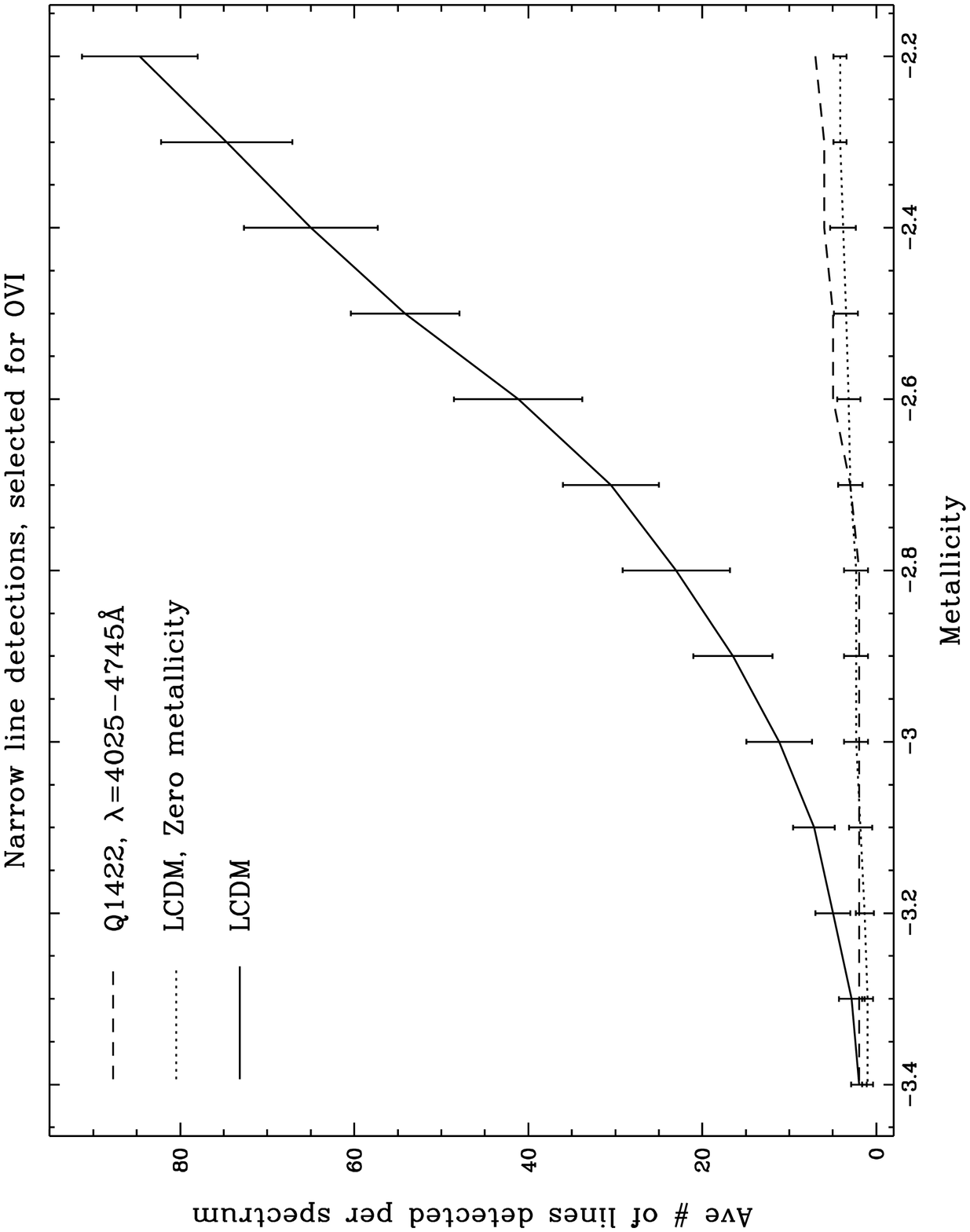}

\clearpage
\plotone{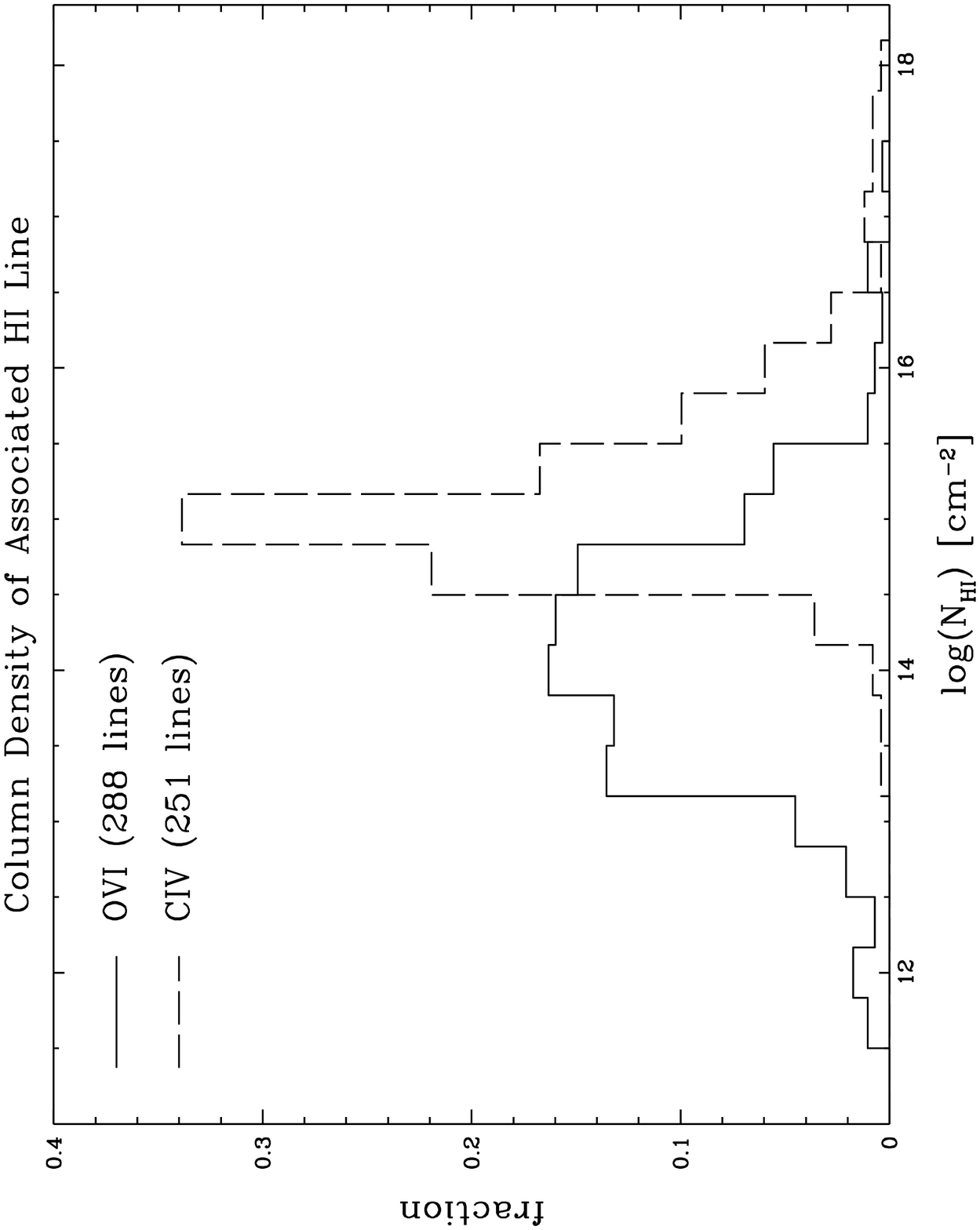}

\clearpage
\plotone{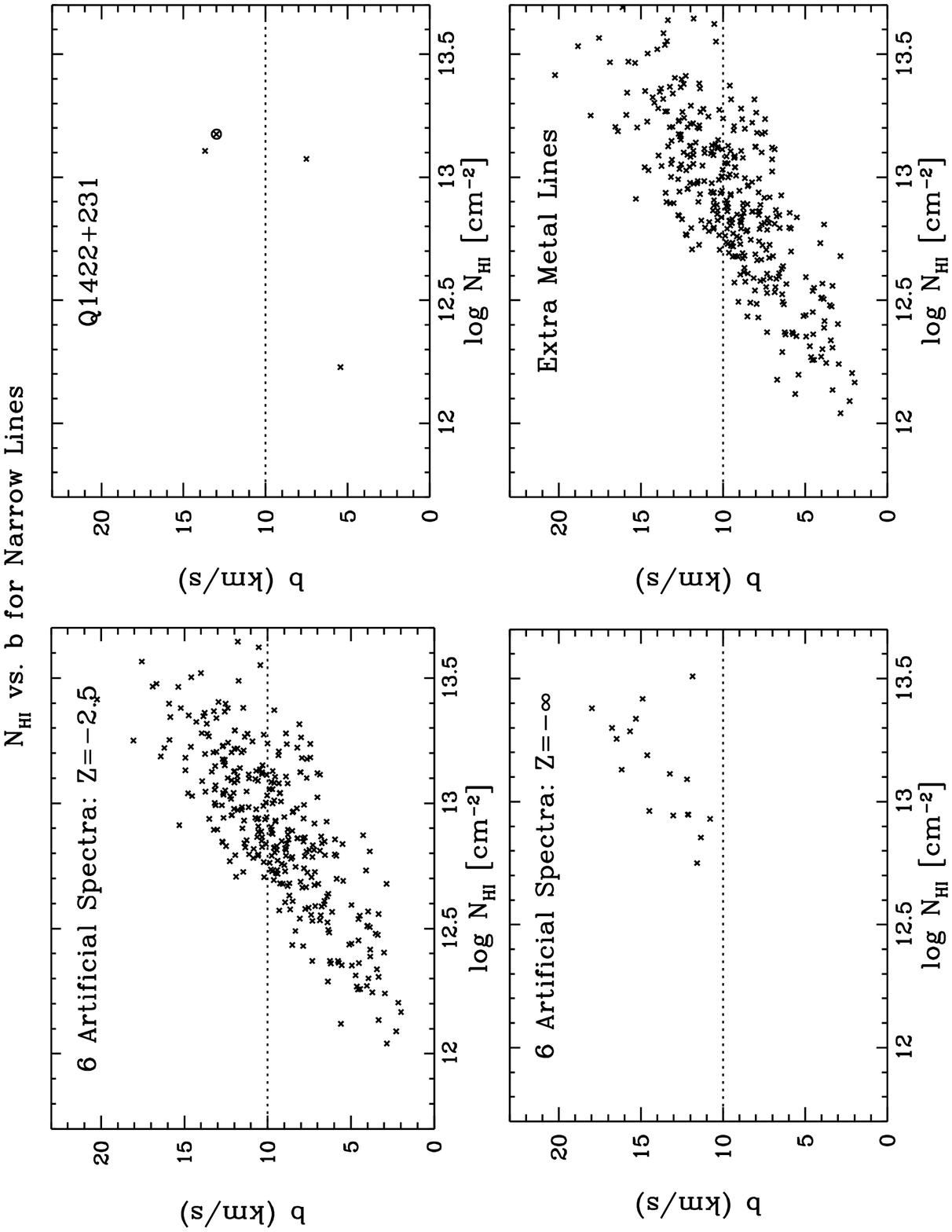}

\clearpage
\plotone{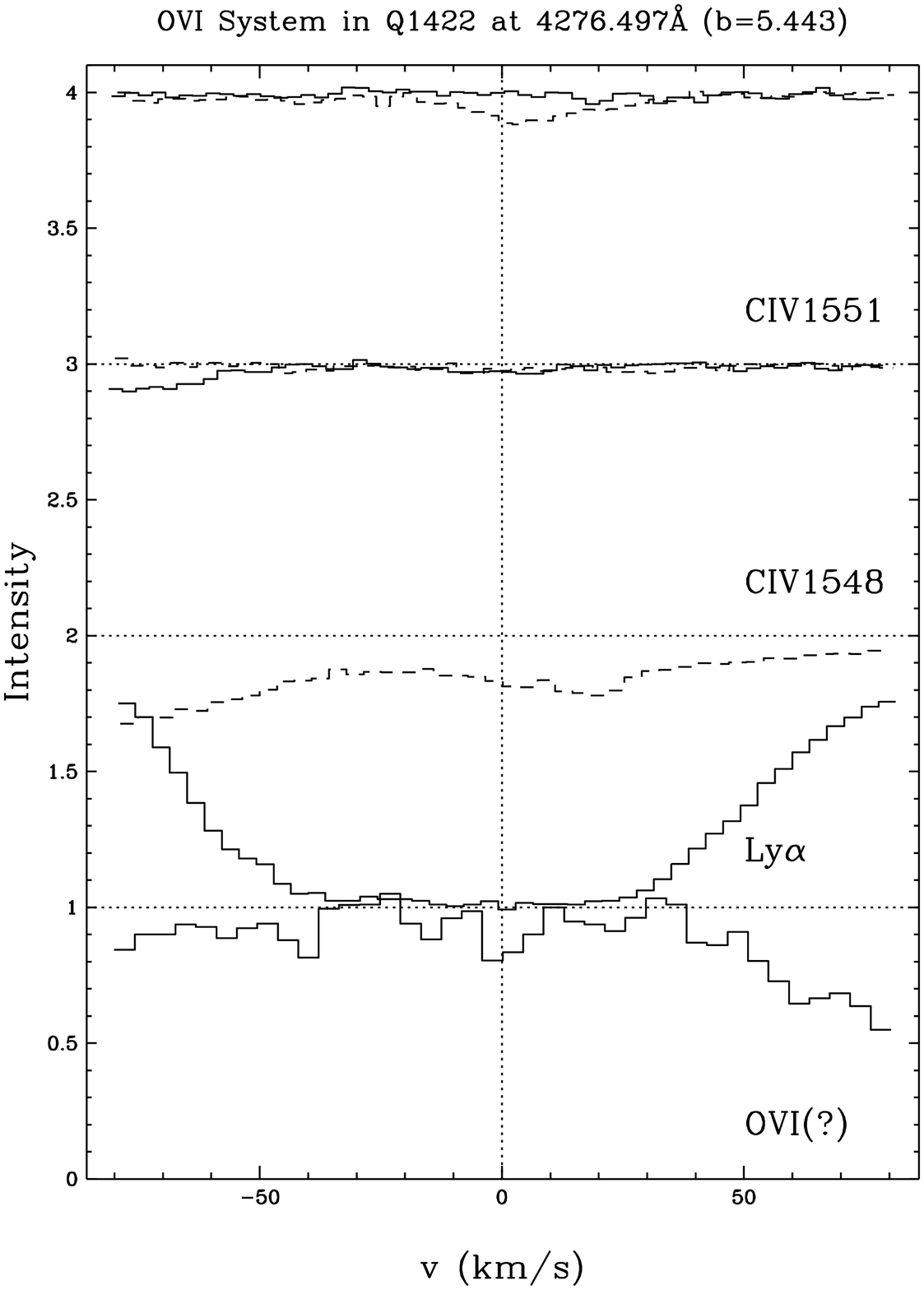}

\clearpage
\plotone{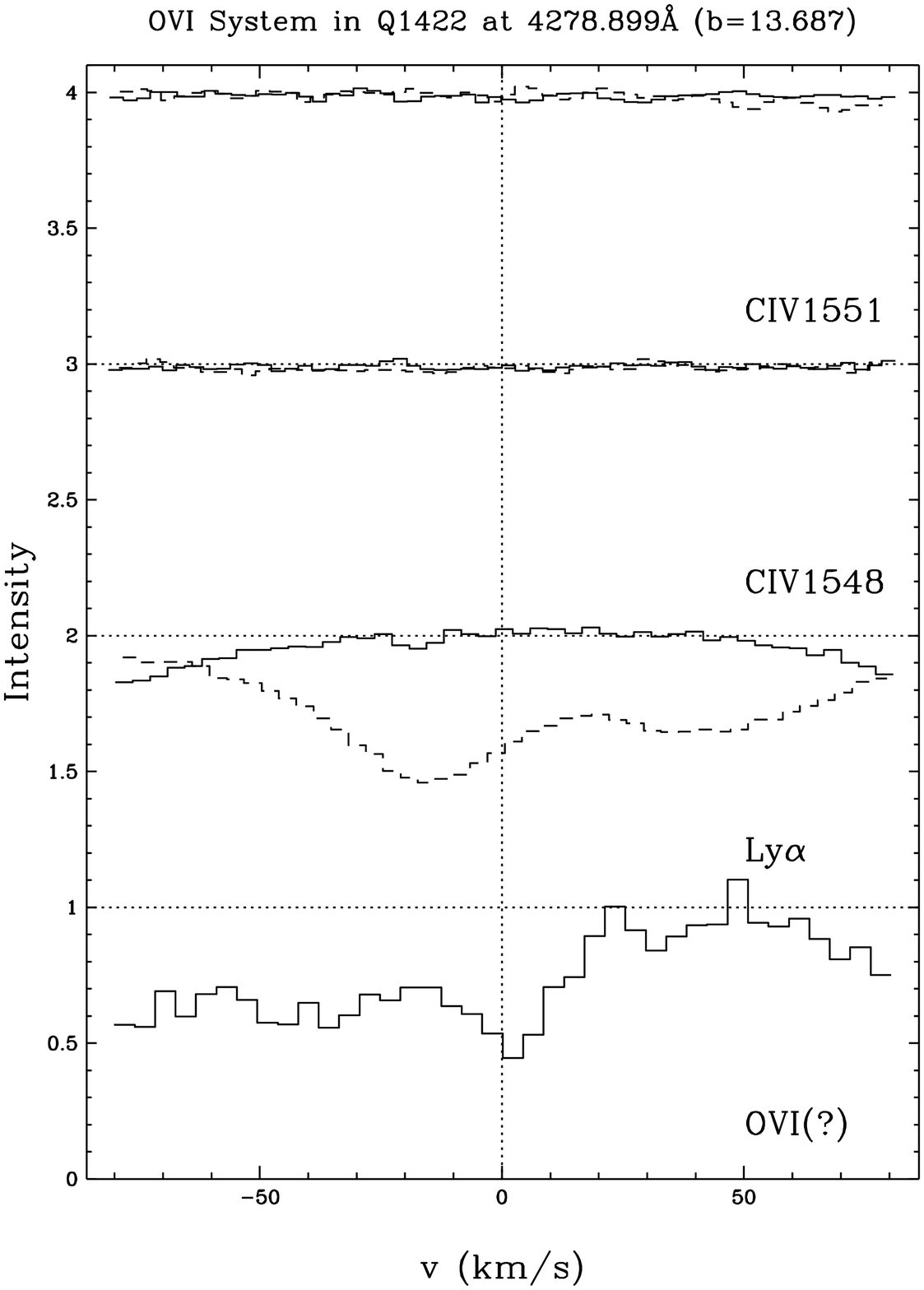}

\clearpage
\plotone{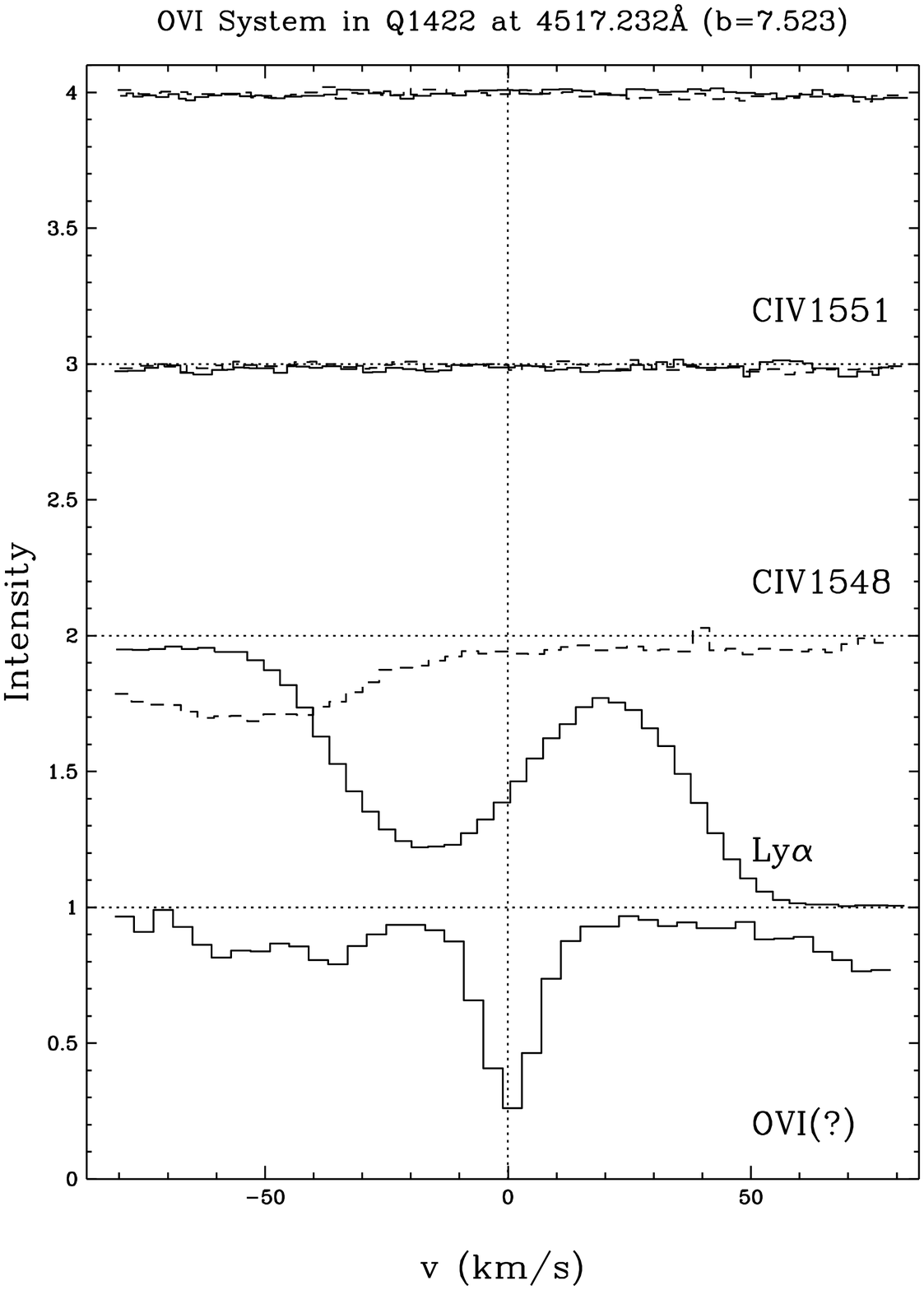}

\clearpage
\plotone{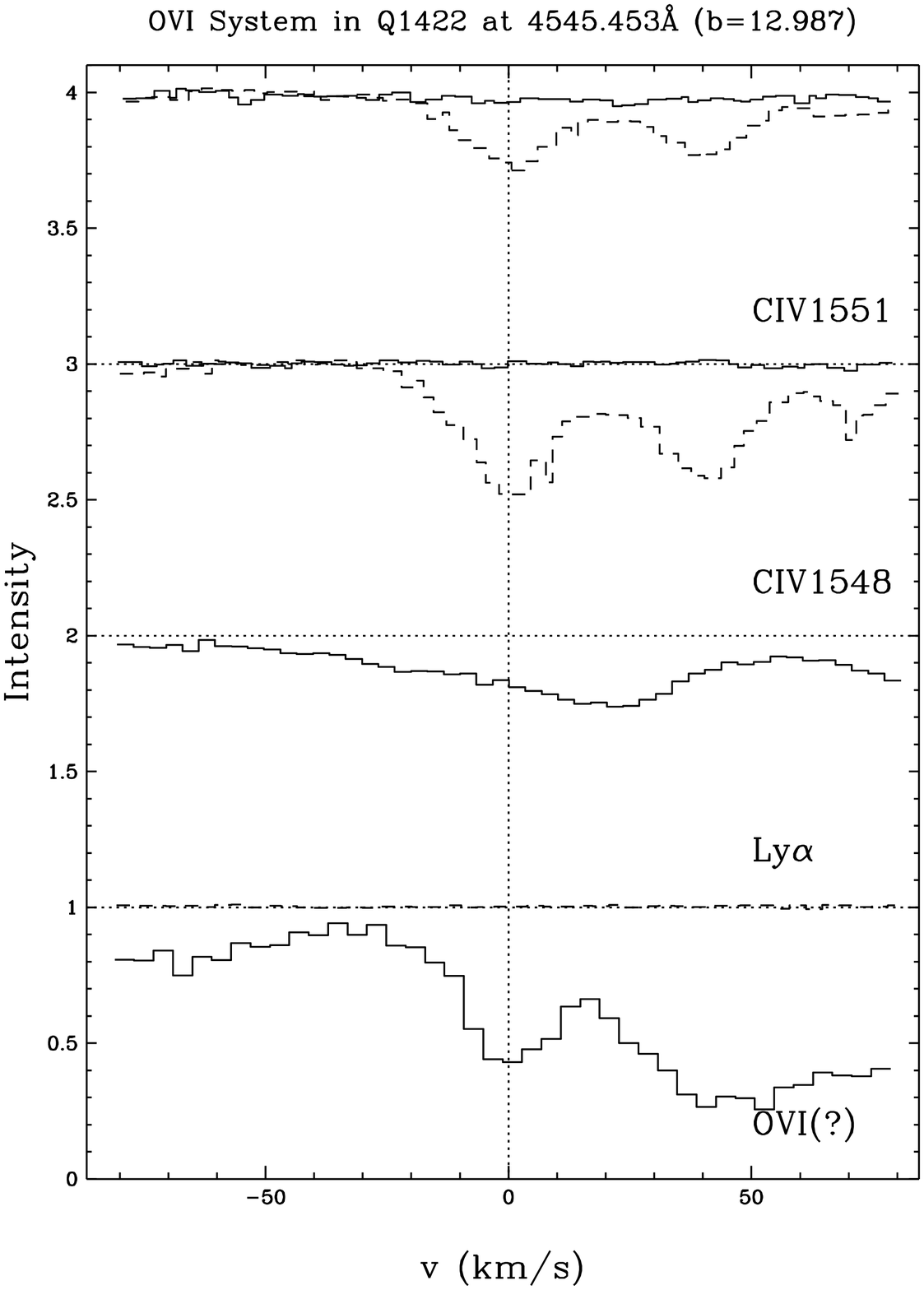}

\clearpage
\plotone{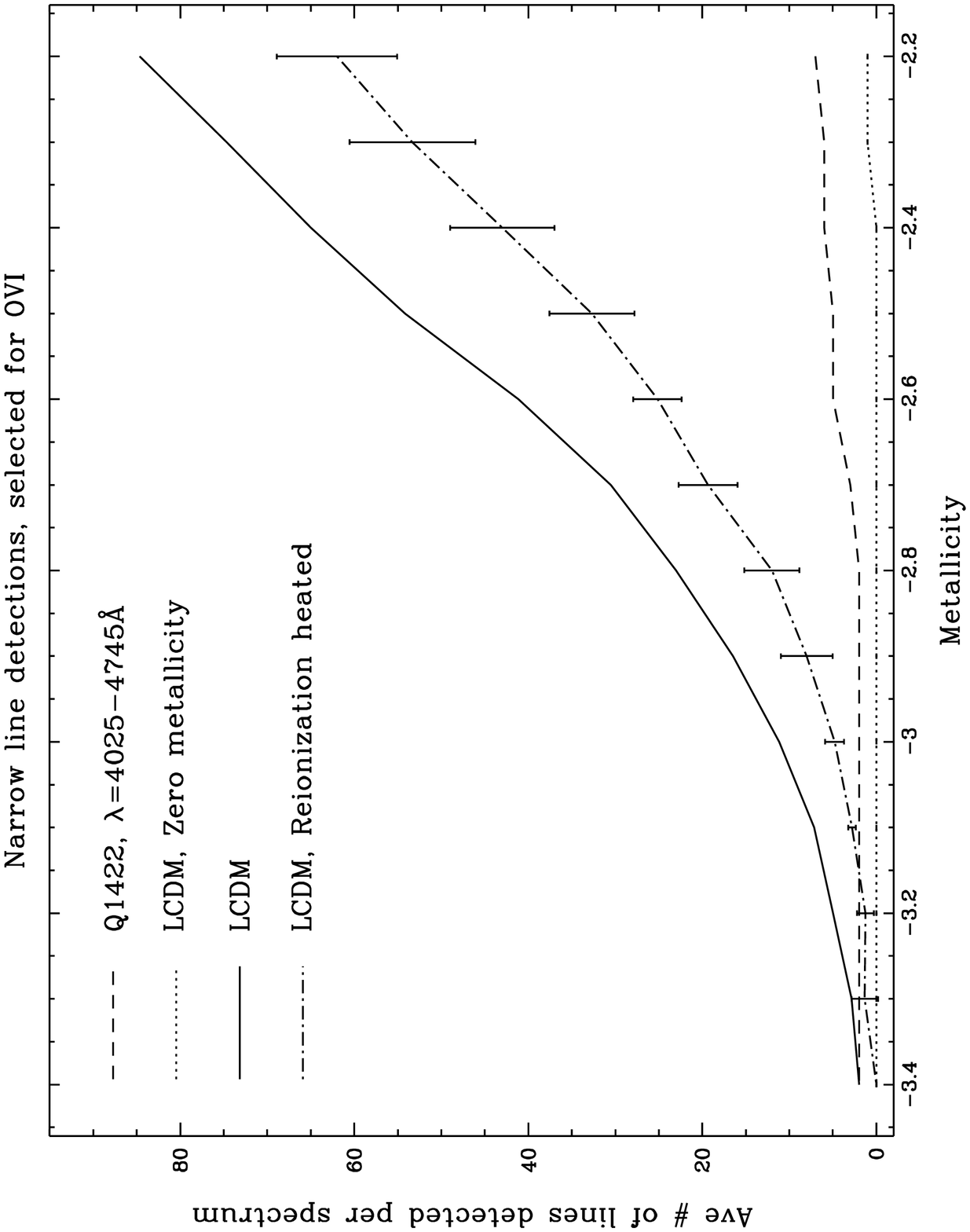}

\clearpage
\plotone{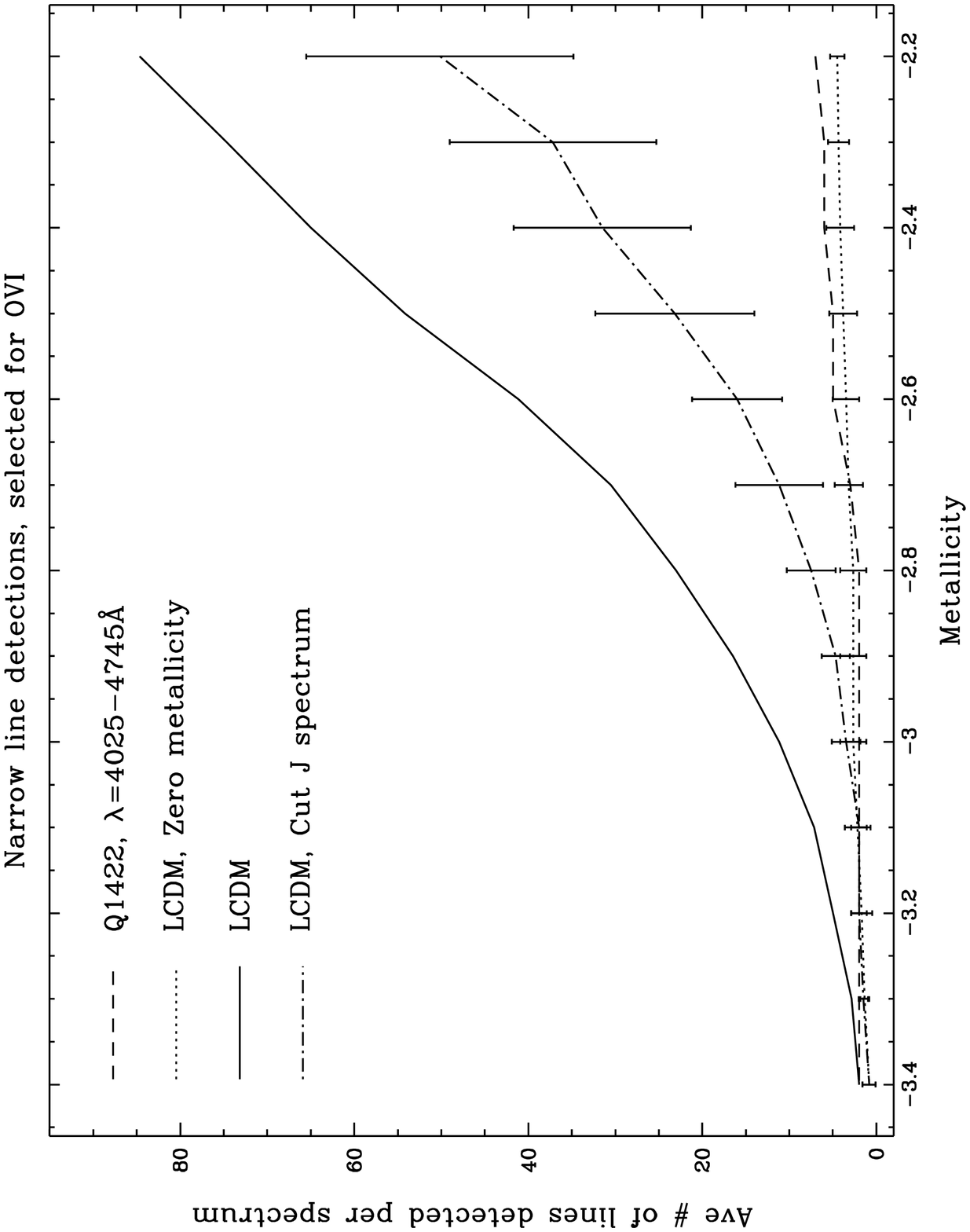}

\clearpage
\plotone{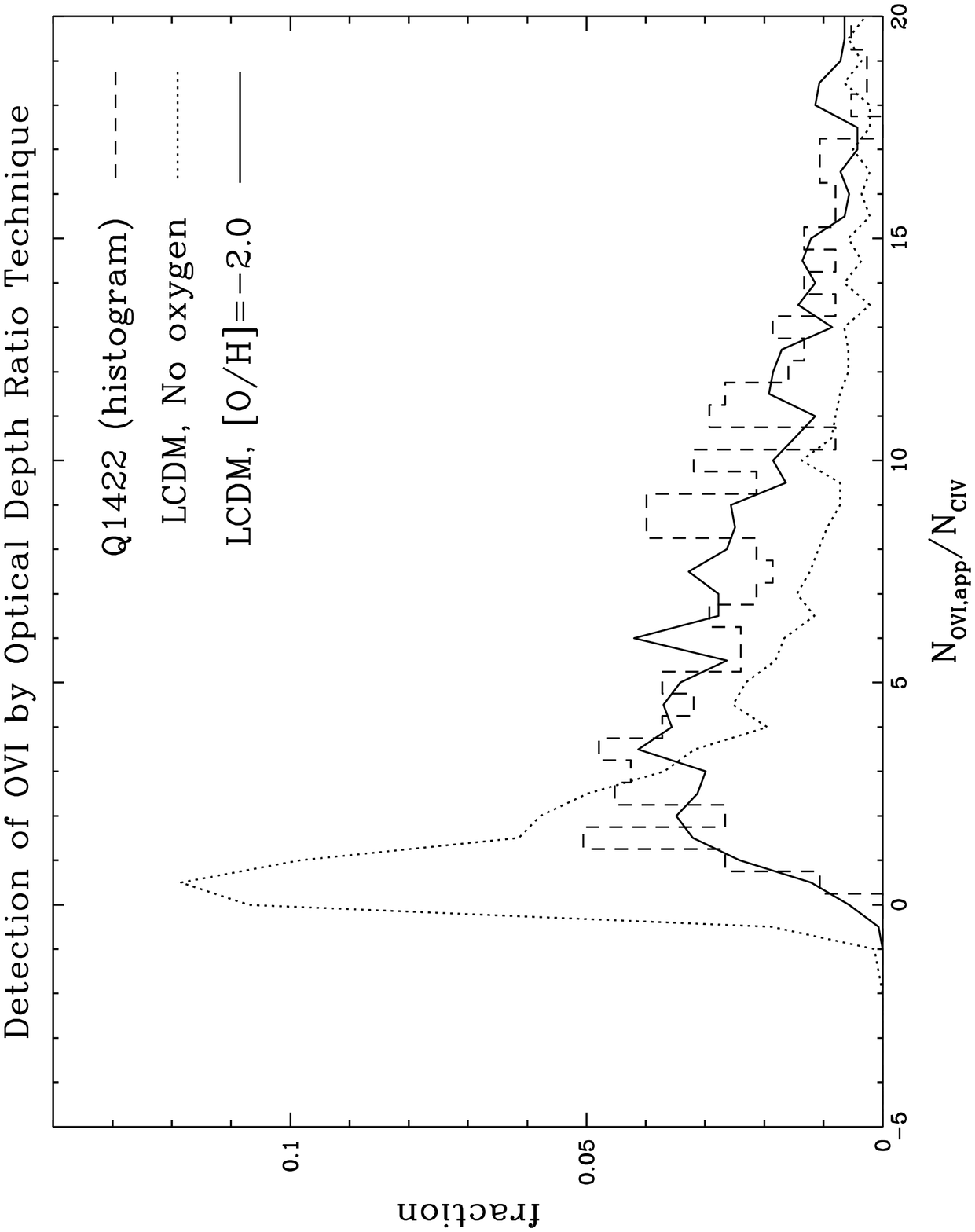}

\clearpage
\plotone{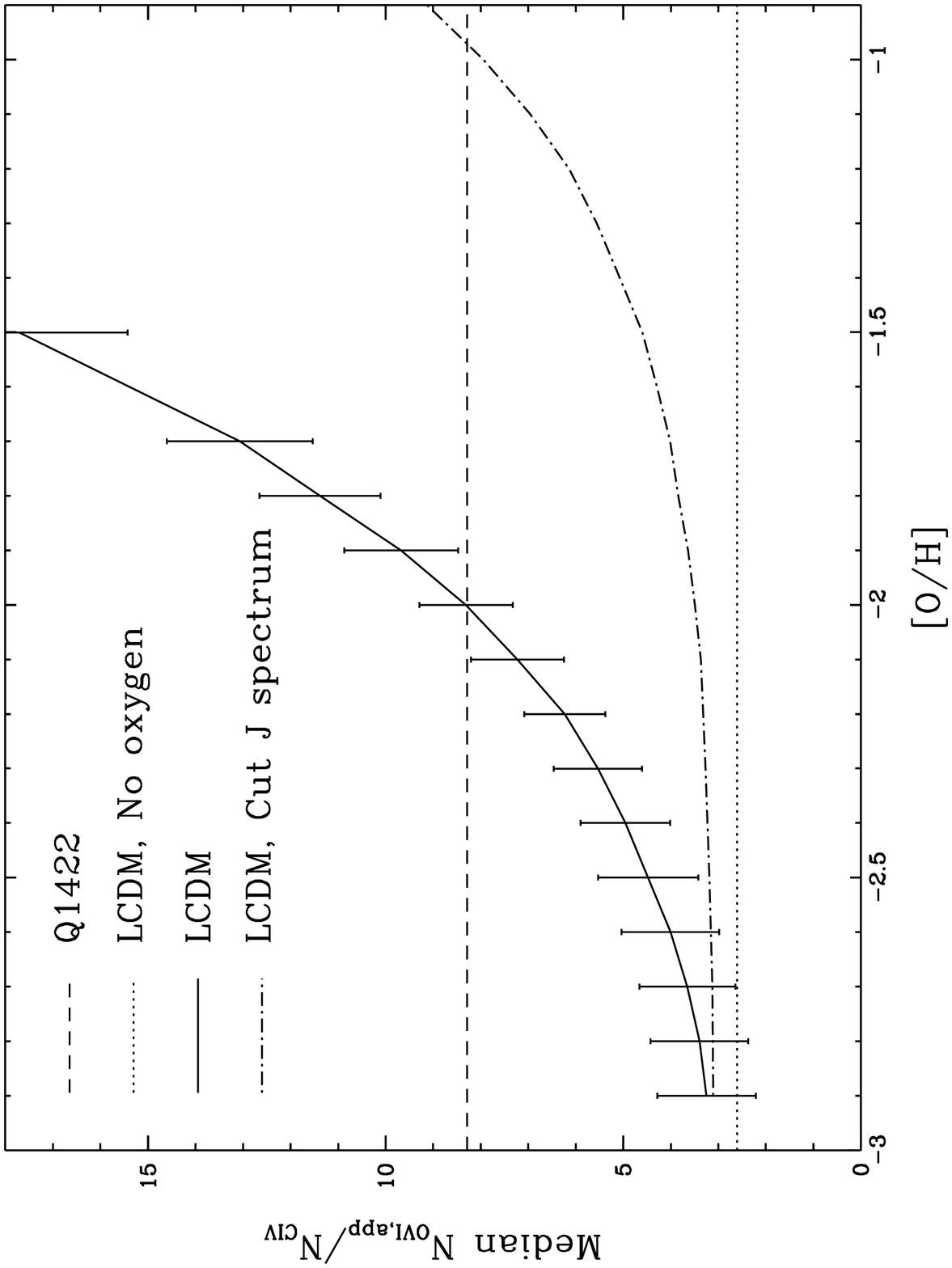}

\end{document}